# Circular City Index: An Open Data analysis to assess the urban circularity preparedness of cities to address the green transition - A study on the Italian municipalities

Alessio Muscillo[1], Simona Re[2], Sergio Gambacorta[3], Giuseppe Ferrara[3], Nicola Tagliafierro[3], Emiliano Borello[3], Alessandro Rubino[4], Angelo Facchini[5*]

[1] *Univesity of Siena, Department of Economics and Statistics, Italy*

[2] *National Research Council - Institute of Geosciences and Earth Resources, Italy*

[3] *Enel X – Innovability Department, Italy*

[4] *University of Bari – Ionian Department, Italy*

[5] *IMT School for Advanced Studies Lucca, Italy*

[*] *Corresponding: angelo.facchini@imtlucca.it*

**Abstract**

We present a circularity transition index based on open data principles and circularity of energy, material, and information. The aim of the Circular City Index is to provide data and a succinct measurement of the attributes related to municipalities performances that can support the definition of green policies at national and local level. We have identified a set of key performance indicators, defined at municipality level, measuring factors that, directly and indirectly, could influence circularity and green transition, with a focus on the green new deal vision embraced by the European Union.

The CCI is tested on a open dataset that collects data covering 100% of the Italian municipalities (7,904). Our results show that the computation of the CCI on a large sample leads to a normal distribution of the index, suggesting disparities both under the territorial point of view and under the point of view of city size. Results provide useful information to practitioner, policy maker and experts from academia alike, to define effective tools able to underpin a careful planning of investments supported by the national recovery and resilience plan recently issued by the Italian government. This may be particularly useful to enhance enabling factors of the green transition that may differ across territories, helping policymakers to promote a smooth and fair transition by fostering the preparedness of municipalities in addressing the challenge.

## 1. Introduction

Strategic and social planning, along with the management of environmental and energy transitions, play a central role in the current and future policy design of regions, cities, and communities around the world (Dall'O, 2020). Sustainability and circularity are fundamental goals of the urban environment evolution and the management of local services assets and infrastructures. As set out by the United Nations, the Sustainable Development Goals are the blueprint for achieving a more sustainable future for humanity, including to "make cities and human settlements inclusive, safe, resilient and sustainable", to "ensure sustainable consumption and production patterns", and to "take urgent action to combat climate change" by integrating related measures into national policies, strategies and planning[1]. Nevertheless, our cities today are far from these goals. Urban areas occupy about 3% of the earth's surface and host more than 50% of the world's population, a figure expected to grow to 68% by 2050 (Lucertini, 2020). Meanwhile, cities are responsible for 80% of the Global Domestic Product and the production of approximately 70% of global $CO_2$ emissions and half of the global waste (UN, 2019; Kennedy et al. 2015, Facchini, 2017; Merino-Saum, 2020).

To address these challenges, in recent years, a large body of studies has focused on measuring urban sustainability through different approaches, methods and metrics, embracing the concepts of sustainability, circularity, urban metabolism and smart cities (Dall'O, 2017; Brillhante, 2018; Sáez, 2020; Sharifi, 2021; Maranghi, 2017). However, the need to identify best practices and help policymakers in addressing processes related to urban and regional planning (Maurya, 2020), remind ourselves that a unique and unambiguous set of strategies to reach these ambitious targets are still lacking. Possible reasons include the wide complexity of aspects and competencies involved, the trade-off between the need for both concise and comprehensive measurements, the diversity of geographical and environmental contexts at the regional and local level, and the different priorities, criticalities and potentials of cities depending on their size and political and socio-economic context (Taylor, 2014). In addition, most studies on urban sustainability and smart cities have so far focused on large cities or urban areas with more than 500,000 people. However, especially in Europe, a significant part of the population (about 40%) lives in non-metropolitan regions or in cities with less than 500,000 inhabitants[2]. This is also true in Italy where, even though most of the Italian population is located in small and medium-sized cities (23% of the Italian population lives in cities with more than 100,000 inhabitants, 47.5% lives in towns with less than 20,000 inhabitants, while 31% lives in towns with less than 10,000 inhabitants), studies regarding urban sustainability mainly involve the largest cities (e.g., Milan, Turin, Rome, Venice, Naples, etc.) or provincial capitals. Thus, the effectiveness of future strategies and policies developed based on a prevailing knowledge of the metabolism of large cities may be poorly suited to the more numerous smaller cities. Taken together, these elements reasonably reduce the availability of tools for planning and monitoring goals and action for ecological and energy transitions, for which cities and local communities alike play a central role.

[1] http://www.un.org/sustainabledevelopment/sustainable-development-goals/

We believe that sustainability policies currently under development in many countries would benefit from scientific support for their definition and integration into appropriate monitoring. To this aim, we propose an index that represents an attempt to measure the development of increasingly sophisticated strategies to effectively implement, manage and monitor the digital, energy and ecological transition of cities and municipalities in Italy. Considering the effort made so far by the scientific community to develop useful indicators to measure the sustainability and circularity of cities (Lucertini, 2020; Merino-Saum, 2020; EIU, 2009), the proposed method is the result of the search for a meeting point between the urgency of shared approaches and tools, the current availability of open data and citizen science. In particular, we consider open data a fundamental resource to support this monitoring task, to stimulate data-driven urban and planning and policy prioritization, and to promote transparency and sharing policy decisions with the scientific community and citizens.

[2] https://ec.europa.eu/eurostat/statistics-explained/index.php?title=Urban_Europe_%E2%80%94_statistics_on_cities,_towns_and_suburbs

The aim of our study is to made available within a single index, the main elements and data that allow the widespread development of analytical tools that permit spatial and econometric synthetic analyses for the identification and monitoring of priorities and critical issues in the current and future sustainability policies of cities and towns at national level, with the main purpose of starting a sustainable and circular path towards an effective and conscious ecological transition. In this study, we put this methodology into practice by addressing an extensive use of the Italian open data landscape, then qualifying ad hoc models to extract relevant KPIs to support the understanding and monitoring of urban circularity, and data-driven policymaking.

As a case study, we collected open data from all the Italian municipalities, gathering data on digitalization, infrastructures, mobility, environment, and waste. The results highlight disparities both at the regional level and at the level of population size, suggesting the need for tools to underpin a careful planning of investment supported by the national recovery and resilience plan recently issued by the Italian government[3]. In particular, as the COVID-19 pandemic recently taught about emergency response, and in light of the European Parliament climate emergency declaration[4] and the ambitious green deal targets, we believe the concept of preparedness should be central in order to make the transition as smooth and fair as possible.

This paper is organized as follows: Section 2 provides a brief overview of the current literature and the main indicators developed; Section 3 presents the foundational methods of the index, and Section 4 describes the application to all Italian municipalities. Results are presented in Section 5 and discussed in Section 6. Conclusions and future directions are presented in Section 7.

## 2. Background and overview of the literature

[3] Next generation Italia https://www.governo.it/sites/governo.it/files/PNRR.pdf

[4] https://www.europarl.europa.eu/news/en/press-room/20191121IPR67110/the-european-parliament-declares-climate-emergency

The global adoption of a sustainable development vision in 1992 marked the beginning of a new era. It was recognized that people's needs and aspirations had to be balanced with healthy ecological and social systems. The pursuit of development, as such, could no longer be justified in economic terms in isolation without the consideration of its broader environmental, social and sustainability impact. Accompanying this dramatic shift in the focus of the development concept, a demand for information and guidance with which to achieve economic, social, and environmental sustainability emerged. One approach to satisfying this demand has been to adopt new indicators of progress that complement and integrate traditional barometers of development that remain mainly anchored to economic variables. Following the early stages of the development of these new and composite indicators (Stiglitz, 2011; OECD, 2011; CEC-COM, 2009), intense production of indicators has emerged in virtually all policy domains, including local communities and cities sustainable development. These indicators typically aggregate a variety of information about a sector and often provide the basis for its ranking, which in turn often allows for comparisons of public policies outcomes. Composite indicators are described as simple to use as a signal, because of their seemingly "simplistic" design and because of the often-unambiguous message they promote, composites are meant to have an impact on high-level policymakers as well as on public and different stakeholders. These indicators have the advantage of being inclusive but the disadvantage of being difficult to decipher and analyse because they strongly condense messages that reflect a very complex reality. In any case, these indicators are of great interest to both policymakers and academics because they provide a set of useful comparators for cities and urban communities. Since in the European Union almost 41.7% of the population lives in cities and 31% in towns, the European Union is solidly committed to making its cities and local communities more sustainable (European Commission, 2010).

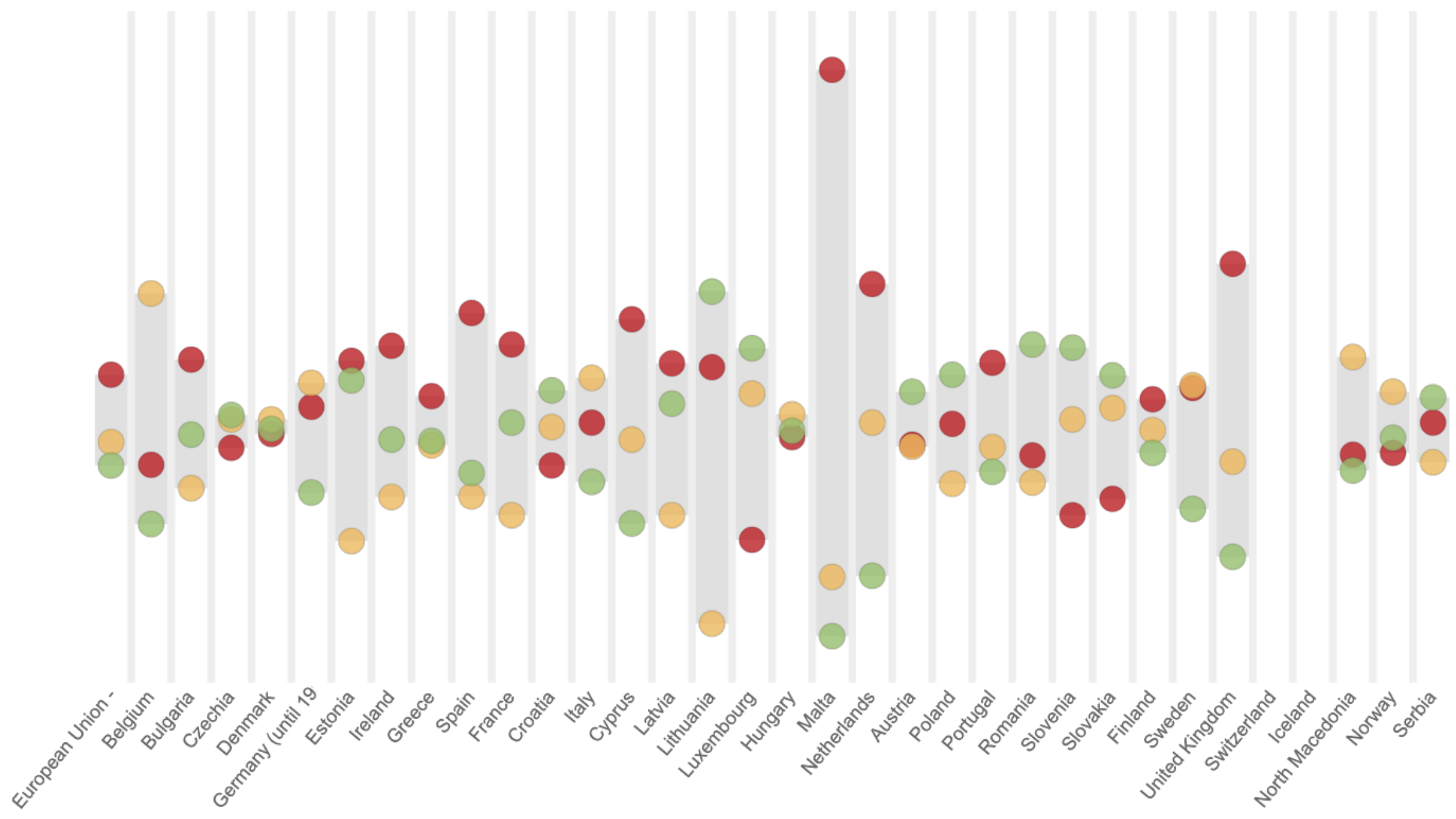

*Figure 1 Distribution of population by the degree of urbanization, Total (Source: Eurostat, 2018)*

To measure the evolving scenario related to sustainability in Europe, several city rankings have been developed that specifically measure the environmental sustainability of European cities. Key examples include the European Green Capital Award, the European Green City Index, and Urban Ecosystem Europe. These rankings contribute to the assessment and development of environmental policy in European cities. However, they also have several shortcomings and important methodological issues. Due to methodological differences, a city may have a high position in one ranking and simultaneously a low position in another ranking. However, objects ranking is inherently problematic, and the data sources used in the European city ranking exercise differ greatly across indicators (see Table 1).

*Table 1 Comparison of the main Indices with the CCI*

| Indicator | Overall index attribute | Number of categories | Number of Indicators |
|---|---|---|---|
| Circular City Index | Focusing on all the municipalities in Italy provide a picture of their sustainable transition | 4 | 17 |
| European Green Capital Award (1) | The overall ranking was a technical combination of 12 separate indicators | NA | 12 |
| European Green City Index (2) | The European Green City Index measured the current environmental performance of major European cities, as well as their commitment to reducing their future environmental impact by way of ongoing initiatives and objective | 8 | 30 |
| Urban Ecosystem Europe (3) | There was no overall ranking attribute. Individual indicators were not aggregated into an overall ranking attribute | 6 | 25 |

(1) Technical Assessment Synopsis Report - European Green Leaf Award 2022, Phrenos (2021)
(2) European Green City Index available here: https://assets.new.siemens.com/siemens/assets/api/uuid:fddc99e7-5907-49aa-92c4-610c0801659e/european-green-city-index.pdf
(3) Maia Da Rocha S, Zulian G, Maes J, Thijssen M. Mapping and assessment of urban ecosystems and their services. EUR 27706. Luxembourg (Luxembourg): Publications Office of the European Union; 2015. JRC100016

While useful and highly informative, they are comparing major European cities in different countries and might provide biased results, as cities are fully immersed in their own social, economic, and political landscape that obviously influence their outcomes and performances across dimensions. On the other hand, while cities are home to the majority of the population in the EU, the composition and distribution of the domestic population vary greatly in each EU country. Also, the relative weight of the city, town and population differs greatly across Europe. Several countries, including Belgium, the Czech Republic, Germany, Hungary, Croatia, Italy, and Lithuania, see a predominance of population located in towns and suburbs and rural areas as opposed to cities. Yet, the results and performances of these areas in terms of their ability to cope with the sustainable transition have yet to be measured and assessed both in the scientific and grey literature.

Aware of the urgency of policy response to the problem of global ecological deterioration and the sustainability requirements of the cities of the future, the aim of our Circular City Index (CCI) is to contribute to fill the above mentioned gaps by assessing urban metabolism variables, thus providing a more comprehensive picture of current municipalities and regions preparedness and strategies to effectively implement, manage and monitor the digital, energy and ecological transition of Italian cities, towns, and rural areas.

### 3. Description of the methodology

The CCI index is structured into four areas: Digitalization (D), Energy, Climate and Resources (ECR), Sustainable Transport and Mobility (M), and Waste and Materials (W) areas. Based on their impact on sustainability and economic circularity, these areas are divided into two groups: a) direct factors (ECR and W) and b) enabling factors (D and M). The latter category includes those factor that can help and facilitate the monitoring and implementation of policies while the former incorporates tools to achieve better results in terms of sustainability and economic circularity.

In turn, each CCI area is composed of several key performance indicators (KPIs), whose importance and weight can vary. In particular, the weight of a KPI increases if it is subject to action by policymakers or if it is subject to political targets and environmental standards set by regulations and laws. However, the application of this methodology to territories – that may vary in terms of populations and ecosystems – should also be considered to increase opportunities for economic and social growth in a sustainable way. As depicted in Figure 2, the KPIs included in the CCI areas can be grouped as follows.

1. Digitalization (D)
    a. Broadband connection
    b. Public websites and digital services accessibility
    c. Adoption of National Digital Enabling Platforms: Public digital identity system (i.e. SPID for Italy) adoption in Public Administration portals & services, Municipalities adhesion to Digital National Registry for Resident Population (i.e., ANPR for Italy)
2. Sustainable Transport and Mobility (M)
    a. Facilities for electric vehicles (e.g., charging stations)
    b. Infrastructures for bicycles and pedestrian areas
    c. Public transport and sustainable mobility
3. Energy, Climate and Resources (ECR)
    a. Resources use and infrastructures
    b. Air quality and emission reduction targets
    c. Renewable and low carbon energy sources
4. Waste and Materials (W)

a. Waste production

b. Waste recycling

c. Electronic waste recycling

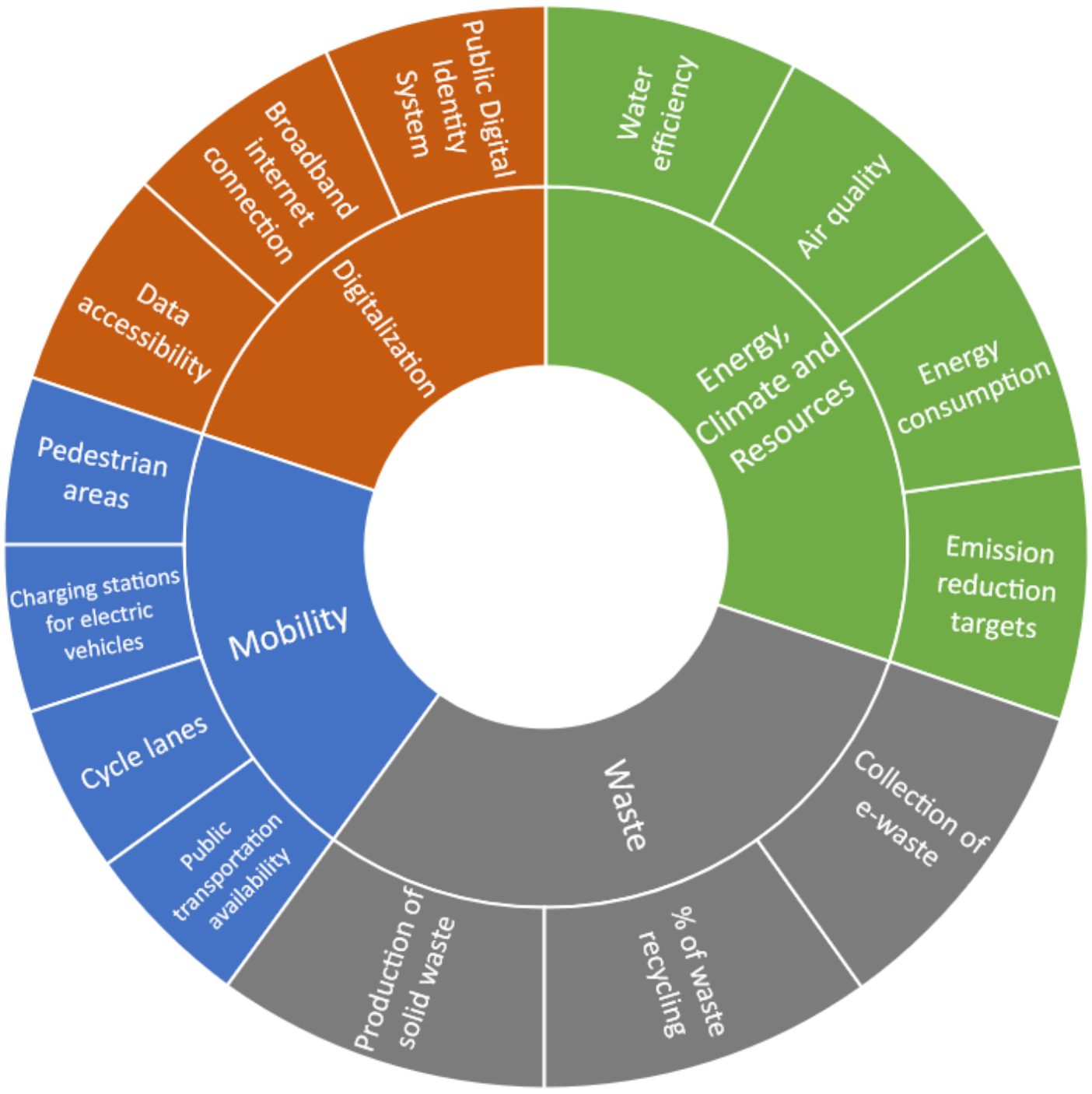

*Figure 2 Schematic diagram of areas and KPIs composing the Circularity City Index (CCI)*

### *3.1 Calculation of the Circularity City Index*

More formally, let us denote a generic area by $A$, the set of KPIs of area $A$ by $KPI(A)$, and the generic KPI in $KPI(A)$ by $k$. Correspondingly, let us denote by $W_A \epsilon [0,1]$ the weight assigned to area $A$, and by $W_k \epsilon [0,1]$ the weight assigned to $k$, which are such that $\sum_{A\epsilon\, Areas} W_A = 1$ and $\sum_{k\,\epsilon\, KPI(A)} W_k = 1$ for every area $A$. Lastly, let us indicate by $S_{kc}$ the results achieved by city $c$ in KPI $k$. Then, the CCI value of city $c$, indicated by $CCI_c$, is a linear combination of weights and KPI values as follows:

$$CCI_c = \sum_{A\epsilon\, Areas}(W_A \sum_{k\,\epsilon\, KPI(A)} W_k \times S_{kc}). \quad (1)$$

The area weights, $W_A$, are derived from the literature discussed in the previous sections, while the KPI weights, $W_k$, are higher if $k$ is a KPI that is subject to policy action. Both types of weights are independent of the city $c$. For a given area $A$, the sub-value obtained by city $c$ in that area is $\sum_{k \in KPI(A)} W_k \times S_{kc}$.

To provide a quicker understanding of the scenario at a higher level, we also applied clustering to the resulting CCI value (and, for further clarity, also to the sub-values of each area D, M, ECR and W) to group the records into the five levels of the Likert scale (Likert, 1932). To achieve this clustering, we used Jenks natural breaks classification method (Jenks, 1967), which is an iterative clustering methodology that aims to reduce in-class variance while maximizing out-class variance. Since this clustering method needs only two parameters (i.e., the dataset itself and the number of desired classes), by using Jenks natural breaks, we ultimately ensure unbiased clustering of records in accordance with the requirements of the entire project (more details in the Supplementary Information section).

**4. Application of the CCI methodology to all the Italian municipalities**

The index illustrated in the previous section is tested on 100% of the Italian municipalities. According to the latest administrative boundaries[5], Italy has a total number of 7,904 municipalities. The opportunity to apply the CCI value to the Italian municipalities, following the economic crisis due to the COVID-19 pandemic, emerged because the Italian government committed to a public investment plan with the aim of triggering and easing a green transition involving the whole Italian territory, with a specific focus on circularity, digitalization and decarbonization of the energy system. Moreover, Italian municipalities are characterized by a wide range of geographic, demographic, and socio-economic characteristics. Table 2Table 2 shows the distribution of municipalities by inhabitants (population). Most of the Italian municipalities (about 90%) are small towns with less than 15,000 inhabitants, which represents around 40% of  the population (about 25M people). On the other hand, only a few municipalities are over 100,000 inhabitants, and a few less of 700 are in the medium-size bracket.

[5] www.istat.it

*Table 2 Number of Italian municipalities and population distribution according to municipality size*

| **City size (population)** | **Number of municipalities** | **Total population** | **Percentage of Population** |
|---|---|---|---|
| < 5k | 5521 | 9.8 M | 16,6% |
| 5k - 15k | 1651 | 14.1 M | 23,8% |
| 15k - 100k | 688 | 21.5 M | 36,3% |
| > 100k | 44 | 13.8 M | 23,3% |
| Total | 7904 | 59.2 M | |

Therefore, it is relevant to understand and to measure how most of the Italian towns will be involved in the green transition, also in consideration of the fact that in the coming years, relevant funds will be invested by the Italian government and significant resources are being allocated to green stimuluspackages.

Starting from the general model described in Section 3, in the next subsection, we describe how the identified KPIs of each CCI area have been implemented considering the Italian municipalities.

### *4.1 Case study and data description*

Among the initiatives implemented in the Next Generation EU plan, the Italian government has recently developed the National Recovery and Resilience Plan (PNRR), which focuses on Italy's green transformation with ambitious policies in the following six areas:

1. Digitization, innovation, competitiveness, culture and tourism.
2. Green revolution and green transition.
3. Infrastructures for sustainable mobility.
4. Education and research.
5. Cohesion and inclusion.
6. Health.

Considering the PNRR, we have computed the CCI considering specific KPIs that are consistent with the plan. In this regard, Table 3 shows the KPIs and the corresponding open data sources used to compose the database. We consider data and KPIs that cover points 1-3, which are, in our opinion, the more related to the concept of urban circularity.

Starting with digitalization, D area, we focused on Public Administration and the availability of broadband connections. On the public administration side, both data on the level of accessibility of public digital service platforms and integration in a public portal of personal digital identity system (SPID) are present in sufficient quality to be used in the study, as is the adhesion of municipalities to the National Digital Registry for Resident Population (ANPR). Data on broadband connection coverage that are collected by the Italian telecommunication authority AGCOM report the share of the population connected to the internet for increasing values of connection speed (ranging from 7Mb/s to over 100Mb/s). Here we have selected the threshold of 30Mb/s, corresponding to a high-quality ADSL2+ connection. This speed limit has been identified by AGCOM as the minimum requirement for a broadband connection and provides the user with sufficient quality for web TV and remote working and education.

For the energy, climate and resources area, namely ECR, we referred to the signing of the Covenant of Mayors and its adoption. Adoption was considered to measure the municipality's level of commitment to European Commission targets, where a simple signature is valued less than the commitment to 2020 or 2030 targets. Air quality is monitored by ISPRA, which is the technical body supporting the Italian government in the management of environmental data flows, and data are available at the municipality level. In this case study, both concentration of $NO_2$ and $PM_{10}$ have been considered, while PM2.5 was discarded because of the lack of data together with its high correlation existing with $PM_{10}$ (Gehrig, 2003; Marcazzan, 2001), and ozone has been discarded because the data was not representative of the entire national territory. To monitor infrastructure efficiency, the lack of data at the municipality level drove our choice towards the water distribution system leakage since data on roads and the electricity distribution system are not available as open data. In fact, the water distribution system is, especially in Italy, subject to regional policies and suffers from several problems mainly related to ageing infrastructures and lack of policy targets that strictly set minimum quality thresholds (the Italian network on average leaks about 35-40%). From an energy perspective, we focused on the self-sufficiency of renewable generation. To calculate self-sufficiency at the municipal level, we referred to public data released by the GSE ("Gestore dei Servizi Energetici"), the Italian public company in charge of monitoring renewable energy production plants. The self-sufficiency KPI, namely ECR3, is computed as follows:

$$ECR3 = \frac{R_{cap}}{H_{cap}} \tag{2}$$

Where $R_{cap}$ is the sum of all the photovoltaic and wind generators capacities present in the municipality area, while $H_{cap}$ is an estimate of the capacity demanded by all the households present in the municipality. Considering an average of 3.3 kW per household, $H_{cap} = N_h * 3.3kW$, the self-sufficiency provides a measure of how the current installed renewable capacity can meet the electricity demand of the households present in the municipality.

In considering indicators for mobility, namely the M area, we decided to focus on sustainable mobility and the availability of local public transportation.

Regarding the EV charging point infrastructure, according to Motus-E, there were 11.834 available charging stations in Italy as of June '21, but since this dataset is not openly available, we opted for the OpenChargeMap project[6] that, by crowdsourcing, collected 8242 charging station locations, i.e. 70% of the total, under an open license with daily updates[7].

Regarding pedestrian areas, bike lanes, and bus stops, features identified by specific tags (see Section S4.2 for more details) were extracted by means of specific queries to the OpenStreetMap database[8].

---

[6] *github.com/openchargemap/ocm-data*

[7] https://www.motus-e.org/wp-content/uploads/2021/07/Istogramma-Storico-Giu-2021.xlsx

[8] ISTAT is currently using OpenStreetMap as an experimental tool for the statistical analysis of road networks (https://www.istat.it/it/archivio/257382)

Finally, for waste (W area), we considered direct factors that measure circularity, including the share of urban solid waste recycled, per capita urban solid waste production, and collection. These data are available in the ISPRA database. Per capita, solid waste production was considered from two different perspectives: recycling and reduction of non-recyclable solid waste disposed in landfills or incinerated.

Table 3 also shows information regarding the value assigned to each KPI and the targets that are considered to calculate the index. The information shown in the table is the type of the KPI, current target, and ideal target. The data are of different types, depending on the nature and objective of the KPI. We count the following types of data:

1) Binary: assuming values 0 (e.g. false, no, etc) or 1 (e.g. true, yes, etc). such type is found when the municipality subscribes or implement a specific policy or is present in a database (e.g. signature of the Covenant of Mayors)

2) Percentage: usually present for material flows or availability of basic services

3) Levels: the KPI takes values in a discrete set of integer numbers (e.g. the level of accessibility of digital public portals or the level of commitment of climate reduction targets)

4) Number: usually expressing a threshold (e.g. the concentration of air pollutants)

To define an objective evaluation, we used the values of the target columns as benchmarks to calculate the value for each KPI (see Supplementary Information for details). The target refers to the current policies implemented to monitor or control the KPIs. Analyzing the targets, we note that most of them are not related to policies currently in force, being ECR4 and ECR5 (air quality), W2 and W3 (waste recycling and e-waste collection), Digitalization (D1 and D2) the KPIs for which a policy is force in Italy. Other targets are only recommended or considered by the Italian government or the EU Commission. These are mainly related to climate reduction targets (ECR1, ECR2) and Mobility (M2, M3, M4) (for a comprehensive view of the target sources, please refer to Section S1 of the Supplementary Information). Other KPIs targets are not linked to specific policy goals, despite being considered in the current EU green deal-related documents.

To set an upper bound for the CCI index, we implemented a set of ideal targets that correspond, for each KPI, to the maximum or most desired value. This choice has been successfully implemented in the literature (see Section 2) and was recently adopted by the Economist Intelligent Unit in the Global liveability index[9].

*Table 3 Definition of the KPI and value levels used in the case study. Targets in bold correspond to policies actually in force; italic targets are recommended from institutions.*

| Area | Area weight | KPI code | KPI weight | KPI definition | Type | Target* | Ideal |
|---|---|---|---|---|---|---|---|
| Digitalization | 0.2 | D1 | 0.3 | Presence in ANPR (public digital service registry for Resident Population) | binary | *yes* | yes |
| | | D2 | 0.3 | Integration of National Digital Identity System (SPID) in Municipalities digital properties (i.e. portals, websites, services) | binary | *yes* | yes |
| | | D3 | 0.3 | % of people with broadband connection (>30Mb/s)[10] | % | *high* | 100% |
| | | D4 | 0.1 | Accessibility of Local government digital properties (i.e. websites, portals) | levels | high | high |
| Energy, Climate and Resources | 0.3 | ECR1 | 0.2 | Covenant of Mayors - Subscription | binary | yes | yes |
| | | ECR2 | 0.2 | Covenant of Mayors - Level of commitment | levels | 2020-30 | 2020-30 |
| | | ECR3 | 0.3 | % local energetic self-sufficiency (from renewables) | % | 55% | 100 |

[9] https://www.eiu.com/n/campaigns/global-liveability-index-2021/

[10] The Italian government considers the limit of 30Mb/s (https://www.ilsole24ore.com/art/banda-larga-casa-casa-ecco-mappa-aggiornata-ministero-AEEUGj7C), while FCC considers 25Mb/s as threshold.

| Area | Area weight | KPI code | KPI weight | KPI definition | Type | Target* | Ideal |
|---|---|---|---|---|---|---|---|
| | | ECR4 | 0.1 | Annual average concentration of PM10 | number | **40µg/m3** | 0 |
| | | ECR5 | 0.1 | Annual average concentration of NOx | number | **40µg/m3** | 0 |
| | | ECR6 | 0.1 | % of water leaks | % | low | 0 |
| Mobility | 0.2 | M1 | 0.2 | Pedestrian areas ($m^2$/100 inhab.) | number | 900 | 900 |
| | | M2 | 0.3 | Charging stations (points/1,000 inhab.) | number | ***1* charging/1,000 inhab** | *1 charging/1,000 inhab* |
| | | M3 | 0.2 | Cycleways (km/100 $km^2$) | number | 100 | 100 |
| | | M4 | 0.3 | Bus stops (number/100 inhab.) | number | 1 | 1 |
| Waste | 0.3 | W1 | 0.4 | Per capita production of solid waste (t/inhab.) | number | low | 0 t |
| | | W2 | 0.4 | % of solid waste recycling | % | **65%** | 100% |
| | | W3 | 0.2 | Collection of e-waste | binary | **yes** | yes |

Table 4 shows the descriptive statistics for the KPIs collected for all the Italian municipalities.

*Table 4 Descriptive statistics of the KPI considered for the case study. Only KPIs with numeric and percentage values are reported.*

| | D3 | ECR3 | ECR4 | ECR5 | ECR6 | M1 | M2 | M4 | M3 | W1 | W2 |
|---|---|---|---|---|---|---|---|---|---|---|---|
| **count** | 7904 | 7904 | 7699 | 7742 | 7904 | 7904 | 7904 | 4514 | 7904 | 7904 | 7904 |
| **mean** | 0.25 | 0.27 | 25.14 | 27.03 | 599.21 | 0.12 | 29.27 | 0.30 | 0.46 | 0.64 | 0.25 |
| **std** | 0.36 | 0.29 | 6.10 | 9.72 | 341.04 | 0.26 | 41.89 | 0.32 | 0.17 | 0.19 | 0.36 |
| **min** | 0.00 | 0.00 | 0.00 | 0.00 | 0.00 | 0.00 | 0.00 | 0.00 | 0.00 | 0.00 | 0.00 |
| **25%** | 0.00 | 0.07 | 20.50 | 21.00 | 265.64 | 0.00 | 0.00 | 0.06 | 0.35 | 0.55 | 0.00 |
| **50%** | 0.00 | 0.15 | 24.00 | 27.50 | 804.30 | 0.00 | 0.00 | 0.15 | 0.43 | 0.69 | 0.00 |
| **75%** | 0.64 | 0.36 | 31.50 | 34.50 | 900.00 | 0.08 | 71.34 | 0.43 | 0.51 | 0.78 | 0.64 |
| **max** | 1.00 | 1.00 | 35.70 | 47.40 | 900.00 | 1.00 | 100.00 | 1.00 | 2.99 | 1.00 | 1.00 |

With regards to the binary and levels KPI, values are observed as follows:

1) D1: 96.7% of municipalities take the value 1

2) D2 86.7% of municipalities take the value 1

3) ECR1 8.7% of municipalities take the value 1

4) W3 90.4% of municipalities take the value 1

5) ECR2 the commitment of the municipalities varies between 0 41.3%; 1 50%; 2 0.5%; 3 1.8%;

| Level of commitment | Percentage of municipalities |
|---|---|
| 0 | 41.3 % |
| 1 | 50 % |
| 2 | 0.5 % |
| 3 | 1.8 % |
| 4 | 6.3% |

## 5. Results

This section shows the results obtained by computing the CCI of all the Italian municipalities. In the first subsection, we show how the the CC! values are distributed and how the single areas determine the final index. Next, we show how the CCI is in relation to the main socio-economic indicators and the results on the robustness check performed on the values of the area weights.

### *5.1 Computation of the CCI in Italy*

Figure 3 shows the distribution of the CCI of Italian municipalities, with the ideal case assuming the value 100 (reported as reference). CCI distribution appears to be normal across the mean value (about 49%), with some slight deviation of the tails from the Gaussian shape. The standard deviation is 9.3, while the maximum and minimum values are 76% and 14%, respectively (refer to Table S1 in the SI section for a full characterization of the distribution). Interestingly, the resulting distribution does not show significant asymmetries or kurtosis, confirming the fact that the CCI calculated on many case studies does not introduce biases that may lead to misinterpretation of the results.

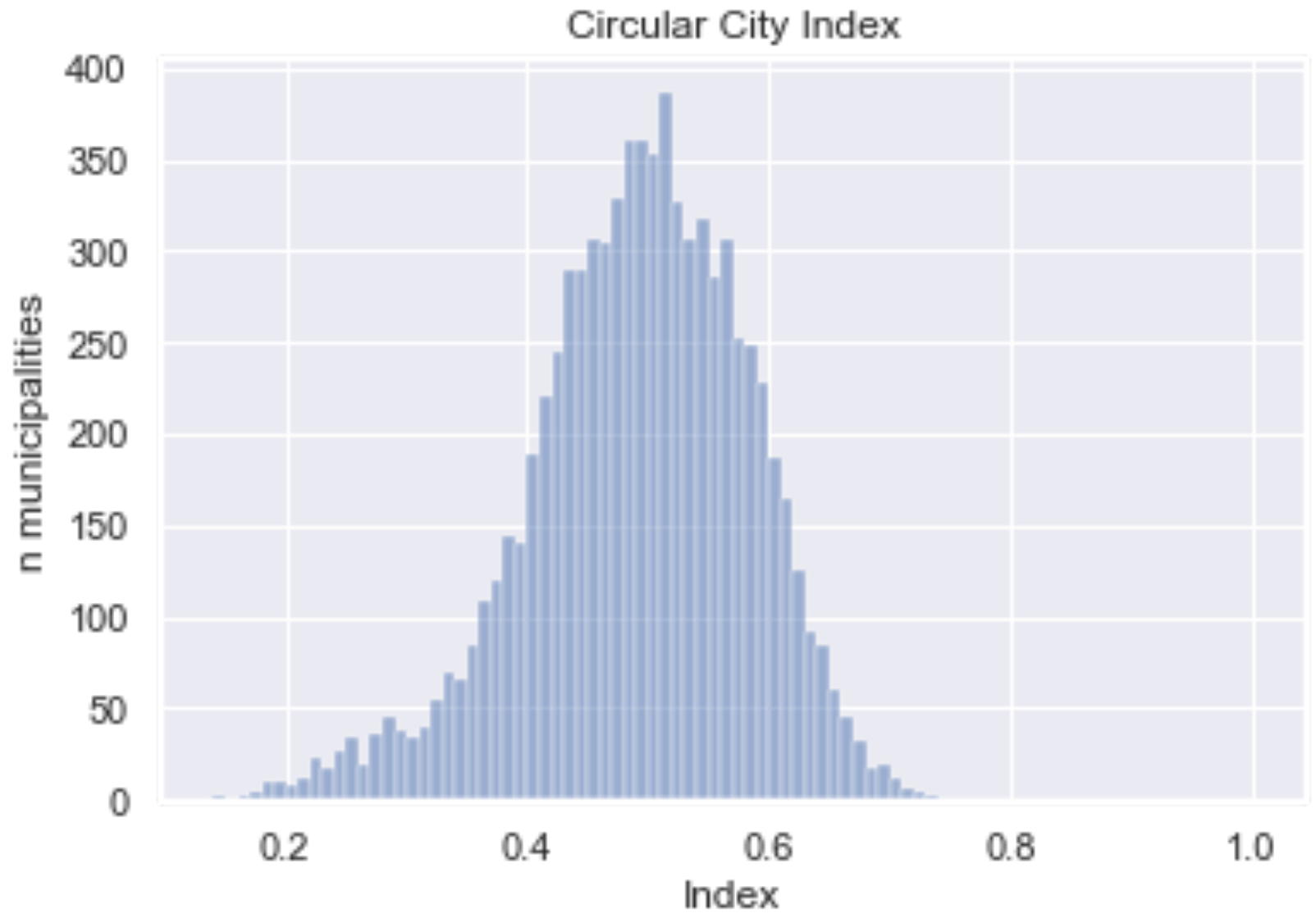

*Figure 3 Distribution of the Circularity City Index across Italian municipalities*

The map depicted in Figure 4 shows how the CCI is distributed in Italy. The map highlights some regional differences: low CCI values are distributed mainly in the southwestern regions, the inner central part (corresponding to the Apennine mountains) and in one region in the southern part. On the other hand, the north-eastern regions generally have a higher CCI value, as well as the regions bordering the Adriatic Sea. Islands generally have a medium to high index.

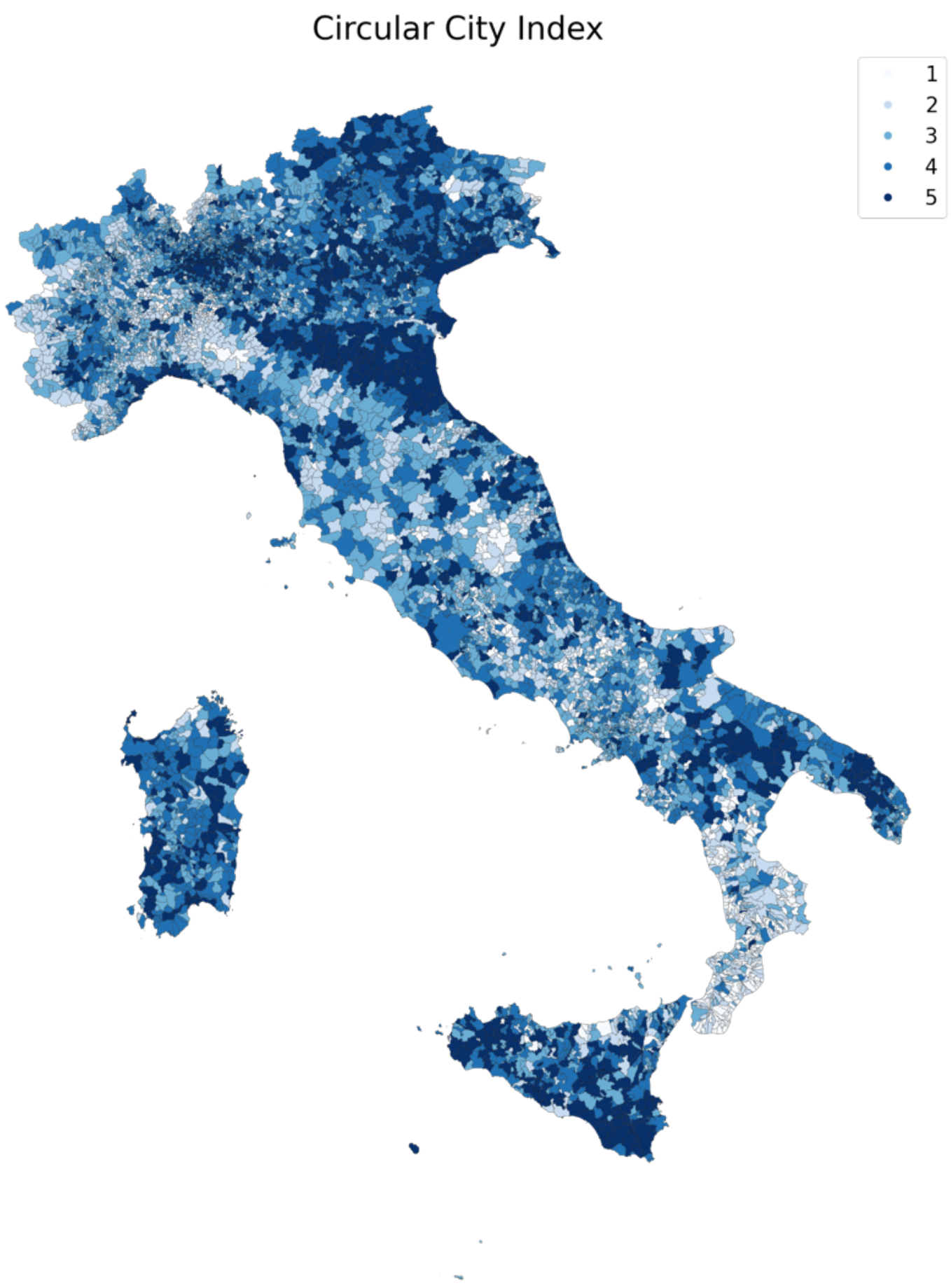

*Figure 4 Geographic distribution of the Circularity City Index in Italy*

To get a better perspective on the composition of CCI, Figure 5 shows the geographic distribution of each of the 4 Area values. Figure 5(a) shows a consistent digital divide between the urban regions of the north, the regions in the central part of Italy, and the south, which, excluding the eastern region and Sicily, generally assume values in the low bracket. The ECR as shown in Figure 5(b) does not confirm the consistent disparity observed in the Digitalisation area, showing that medium to high values are homogeneously distributed across the regions. The largest low-scoring area is in the north-western part of the country. It is worth noting that the islands' municipalities generally show high values, mainly because renewable energy sources have been encouraged by regional policies in recent years. Figure 5(c) depicts the Mobility area value, showing the widest disparity between northern and southern Italy where, except for dense urban areas, the value is generally in the lower bracket. An additional element that lowers the mobility value is related to the lack of light mobility infrastructures in southern municipalities (e.g., lack of bike lanes). This difference can be attributed to different factors: the first can be found in the data source[11], since the main source of open mobility data is OpenStreet Map, whose sampling might not be uniformly distributed.; other factors might be related to different socio-economic conditions, which drive different policies on sustainable mobility (e.g., creation of bike lanes or distribution of charging points). Lack of public mobility may also be a driver for the mobility divide. In the end, Figure 5(d), which shows the Waste area  points out a generally medium to a high value, with the exception of some mountainous regions. Small municipalities in the south generally show higher due to their low waste per capita and high recycling rates. E-waste is not always collected, further reducing the area value of these municipalities.

The maps shown in Figure 4 and Figure 5 revealed some disparities that are worth to be investigated at a finer scale than regional. We now consider the distribution of CCI by different city sizes, with the aim of understanding whether the disparities that emerged are also found between larger and smaller urban centres.

[11] See Table S1-1 and in Section S2.1,

(a)

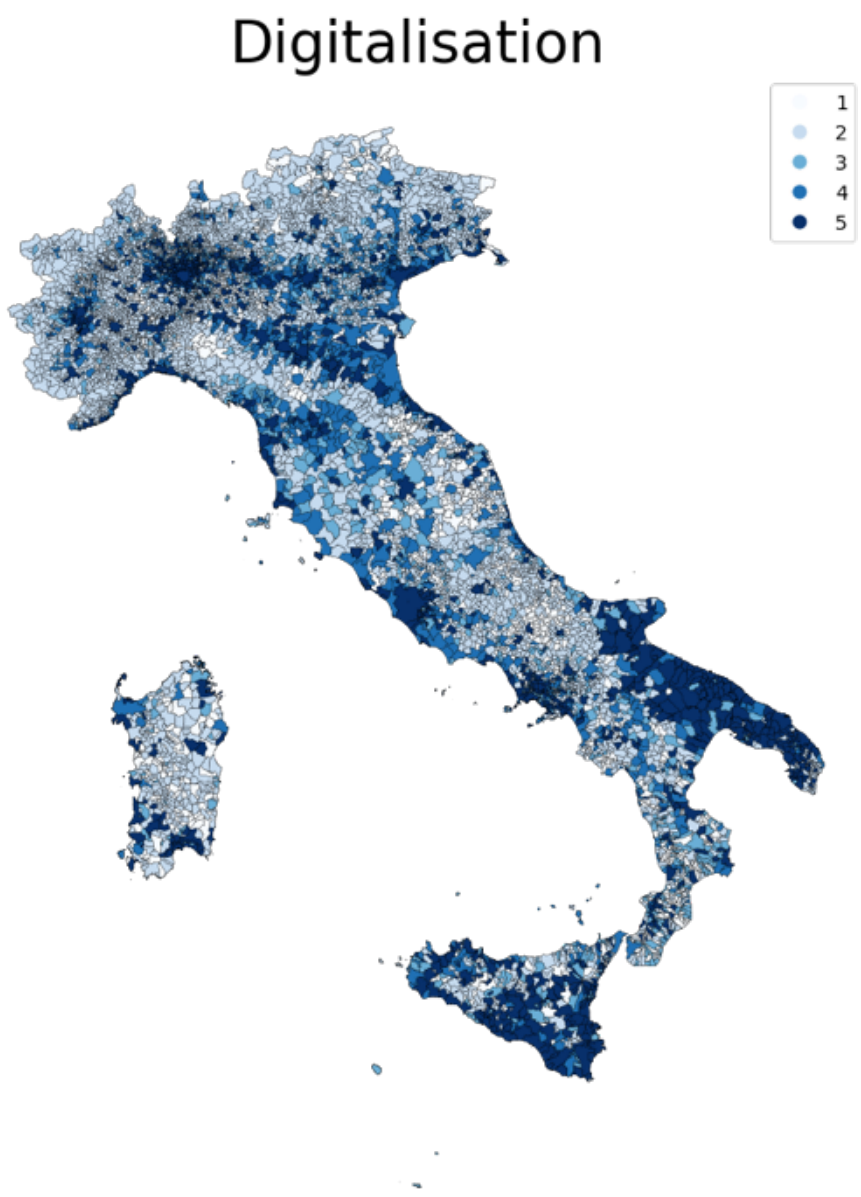

(b)

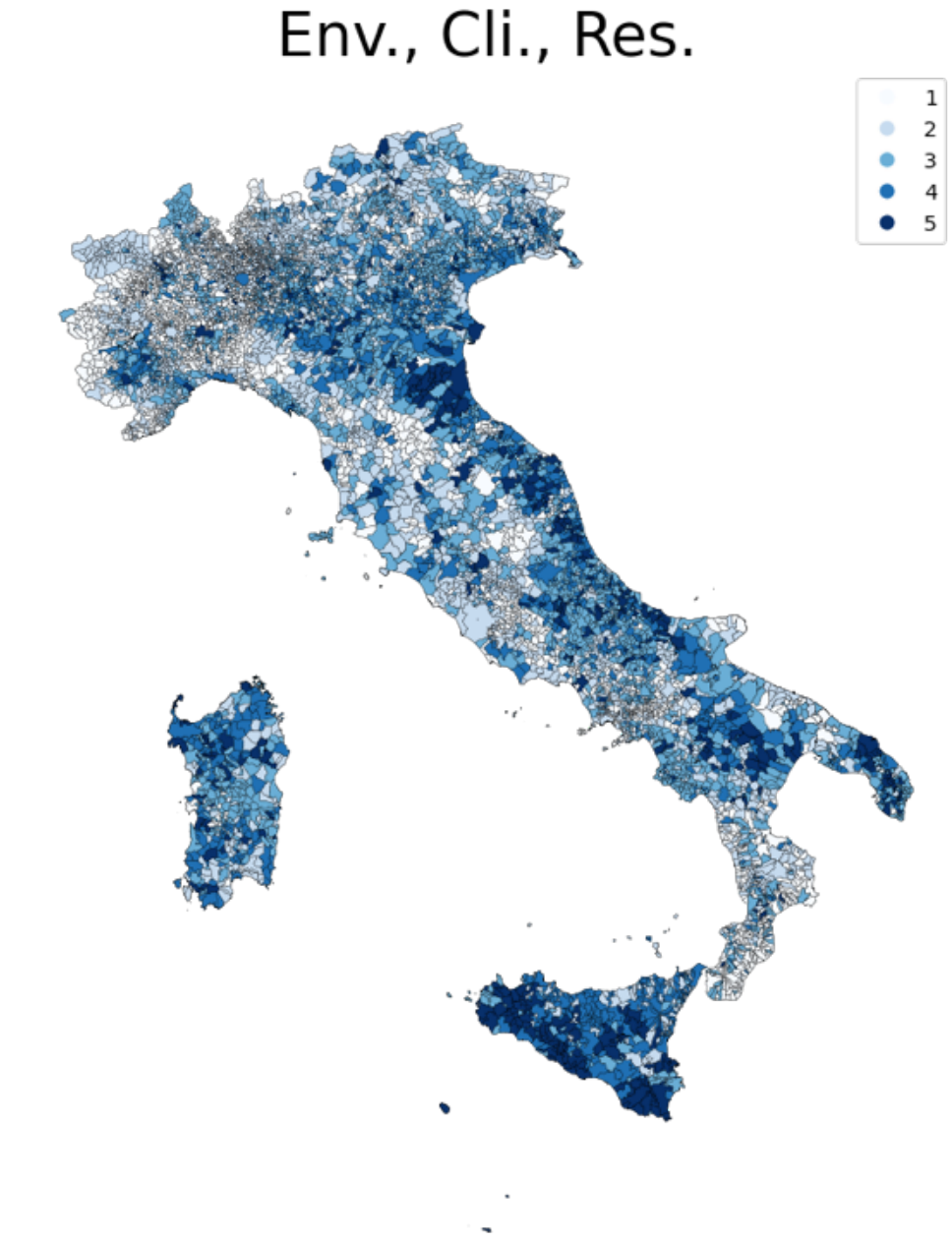

(c)

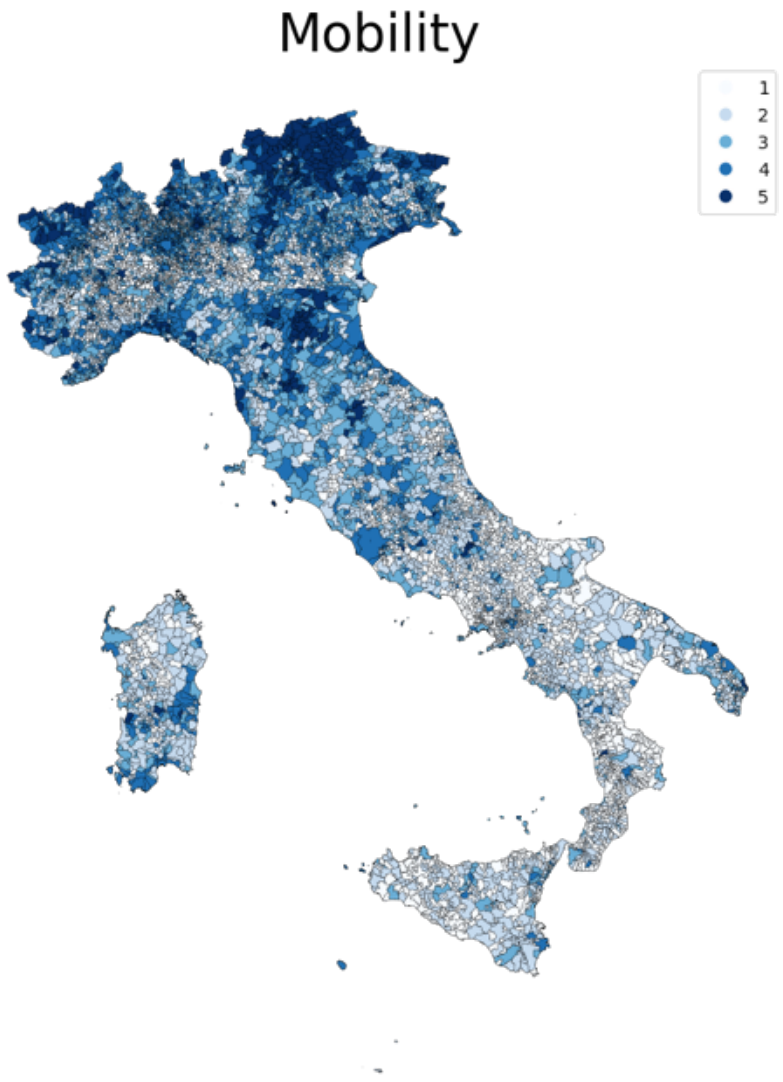

(d)

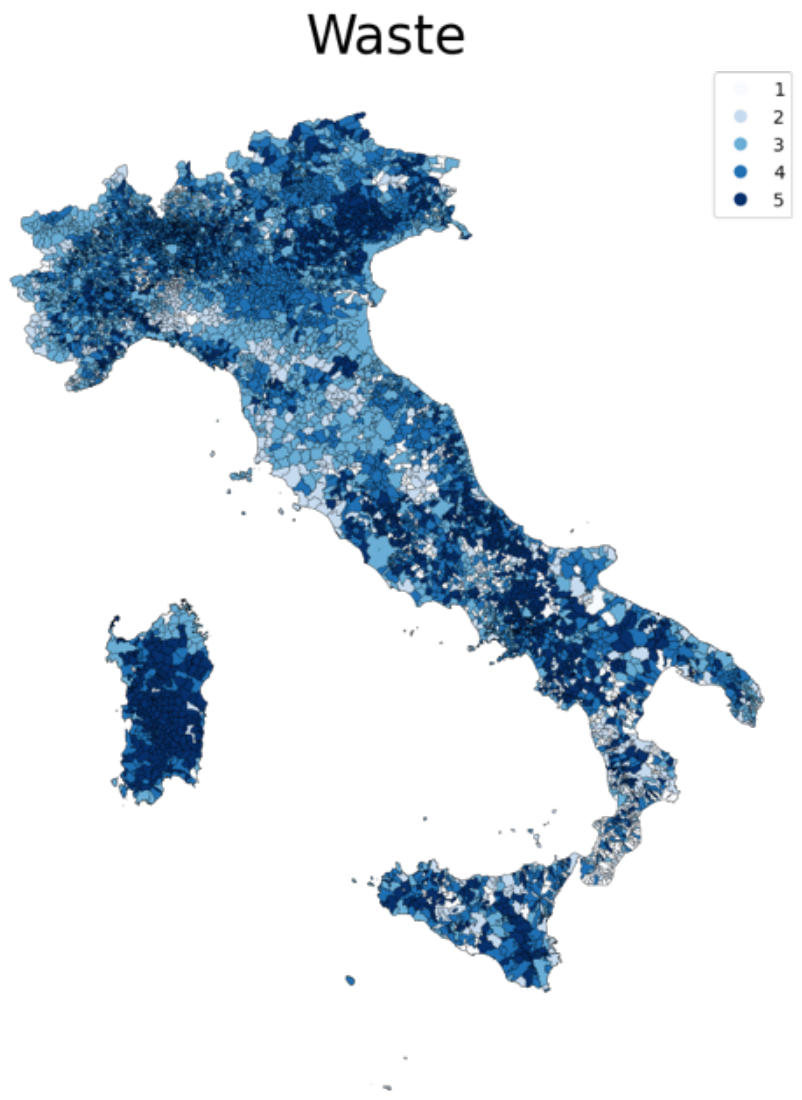

*Figure 5 Distribution of the area values in Italy. (a) Digitalization; (b) ECR; (c) Mobility; (d) Waste*

Figure 6 shows the distribution of the CCI according to the size of the municipality. The average index increases with the size of the municipality: those with more than 100,000 inhabitants are on average about 0.58, while small municipalities are on average 15 points less. Medium-sized municipalities (5k-15k and 15k-100k inhabitants) are distributed in the 50-60% range. Analysing the distribution of area values, we note that the distribution corresponding to the different sizes of city seems to span over the same range for all dimensions, except for the Digitalisation area, which shows a relevant peak for both small (centred between 0.1 and 0.15) and medium-large cities (centred between 0.15 and 0.2).

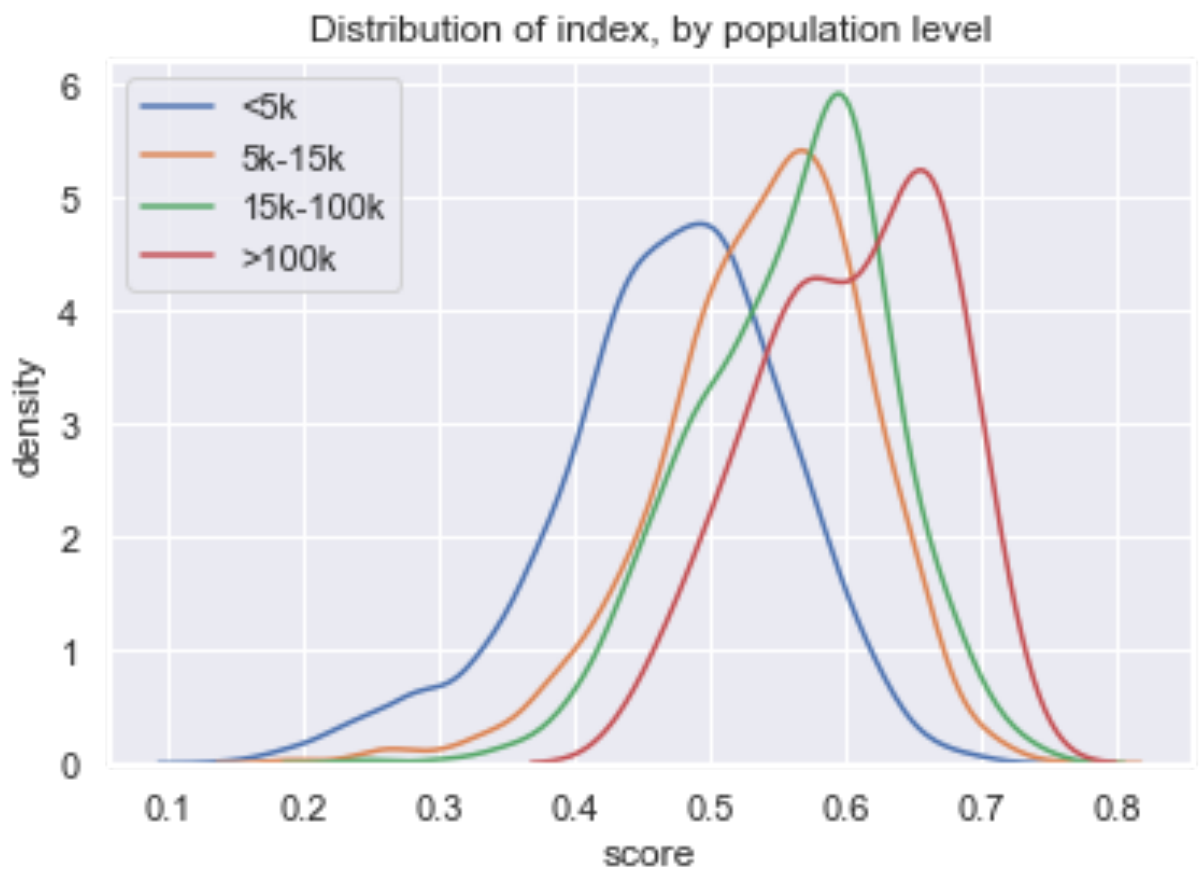

*Figure 6 Distribution of the CCI with respect to the municipality size*

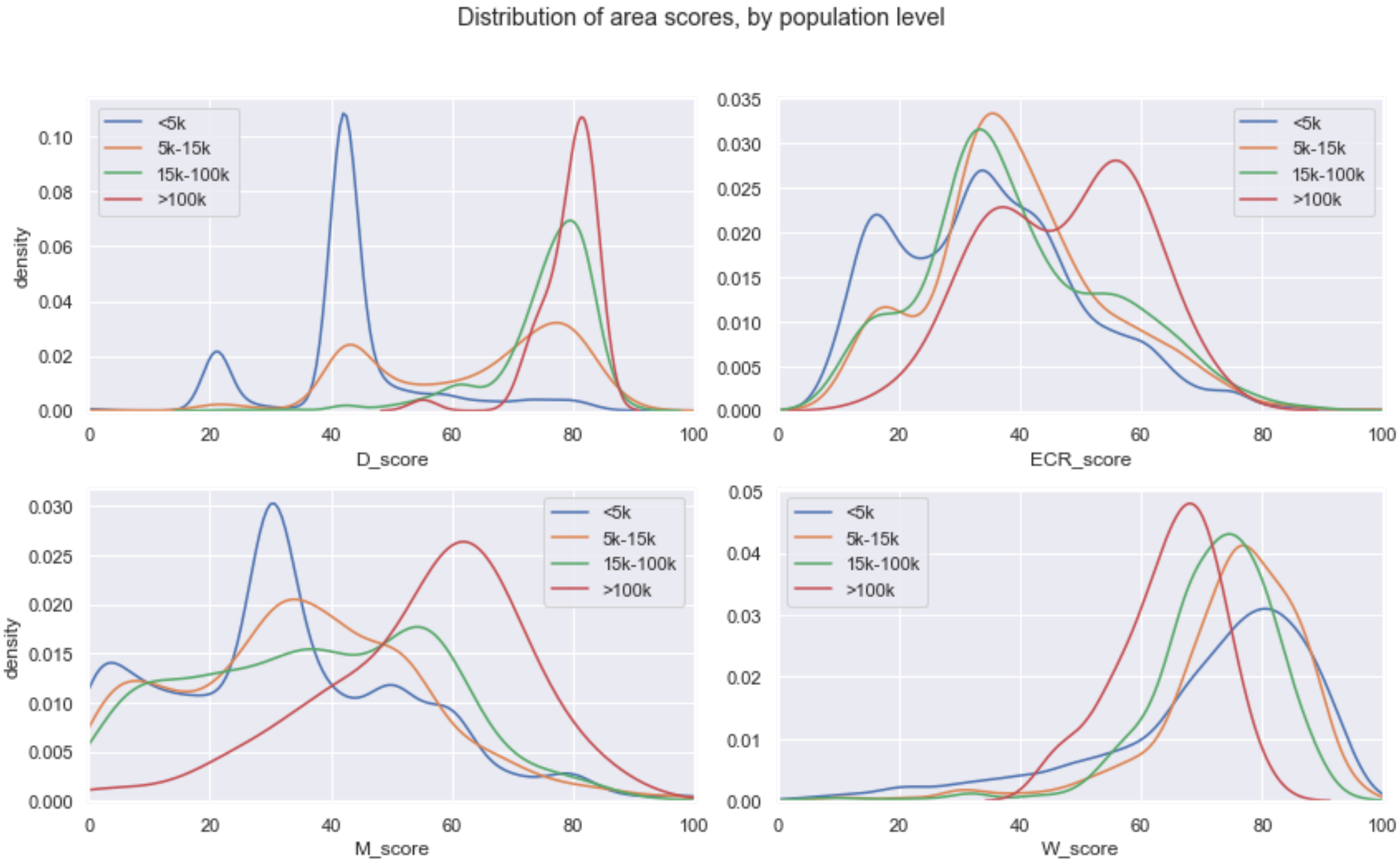

Figure 7 shows the distribution of area values by municipality size. The largest gap is found in the Digitalisation, with a consistent difference between large and medium-large municipalities (80%) and small municipalities (40%), while better uniformity is found in the ECR and Mobility areas, where generally larger municipalities show higher values. An exception to this is found in the Waste area, where small municipalities show higher values with respect to the larger ones.

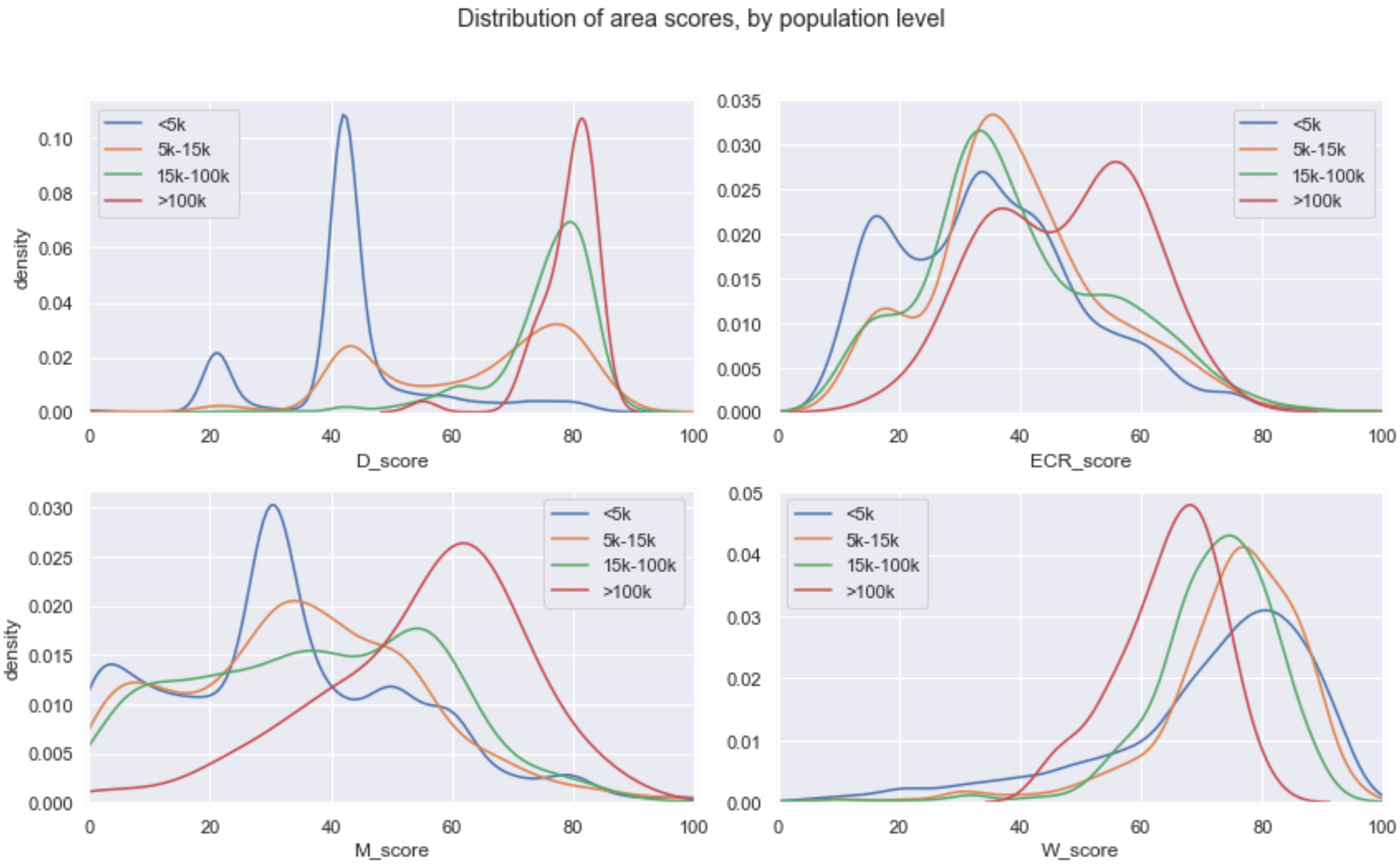

*Figure 7 Distribution of area values for different municipality sizes*

We now consider the number of citizens living in municipalities by CCI percentile. Table 5 shows that most of the Italian population lives in areas with medium and high CCI, while about 13M people live in the low and medium-low bracket. The table also shows that the lowest CCI values are more likely to be found in small municipalities (average size up to 4.5k inhabitants).

*Table 5 Number of inhabitants living in municipalities with low-medium, medium-high, and high CCI values.*

| Quartile | Inhabitants | Quartile range | Average size |
|---|---|---|---|
| 1st | 4.6M | 0-43.5 | 2.3k |
| 2nd | 8.91M | 43.5-49.7 | 4.5k |
| 3rd | 13.6M | 49.7-55.7 | 6.8k |
| 4th | 32M | 55.7-100 | 16.2k |

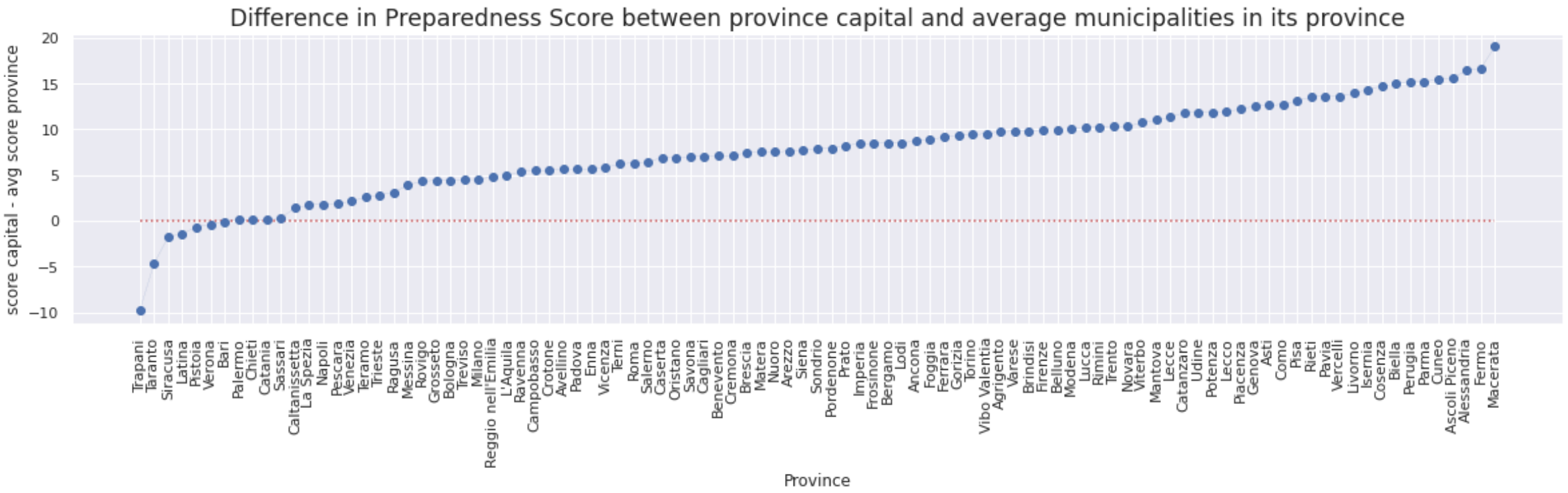

*Figure 8 CCI difference between province capital and surrounding municipalities.*

In Figure 8 we also investigate the CCI differences between the provincial capital and the surrounding municipalities of each Italian province. The vertical axis of Figure 8 shows the difference between the value of the provincial capital ("Capoluogo di Provincia") and the average values of all the other municipalities in the same province. Excluding a few cases, in general, capitals show values higher than the other municipalities, and in many cases the difference exceeds 10 points. This suggests that infrastructural development (e.g., digitalization, mobility, energy) and environmental standards have been better developed in the capital municipalities. Looking at individual areas, this pattern is found in all areas except waste, where the capital cities show lower values. This is likely because scaling up waste collection and differentiation in the city is more complicated.

*5.2 Relation with key socio-economic indicators*

We now consider the relation of the CCI with other socio-economic variables with the aim to highlight possible correlations with other variables. In this paper, we consider population and income (both total and per capita) and their corresponding variation in recent years. More precisely, we consider for population a 10-year time span (2011-2021), while for the income we refer to the years 2014-2018 (time span has been restricted to match the availability of data in the ISTAT database).

Figure 9(a) shows the correlation matrix of the main area values considered in this case study. We notice that the area values and CCI generally show low or near-zero correlation values, further confirming that the proposed index could be able to capture the preparedness of the Italian municipalities by going beyond the main socio-economic drivers and their temporal variation over a period of several years. Going into more detail, when considering population, the CCI shows a weak a significative correlation (0.45) with the logarithm of population. This value is probably due to the value of 0.62 found between the Digitalisation and the population. The correlation values drop significantly when considering the change in population over the last ten years, showing values between 0.29 and 0.01.

Similar results are found when considering per capita income, the correlation values drop significantly, assuming values between -0.13 and 0.44.

These correlation values also suggest that both high and low preparedness do not appear to be correlated with income and population dynamics of the municipalities. For example, municipalities that have shown declines in both income and population in recent years do not necessarily have low CCI values.

(a)

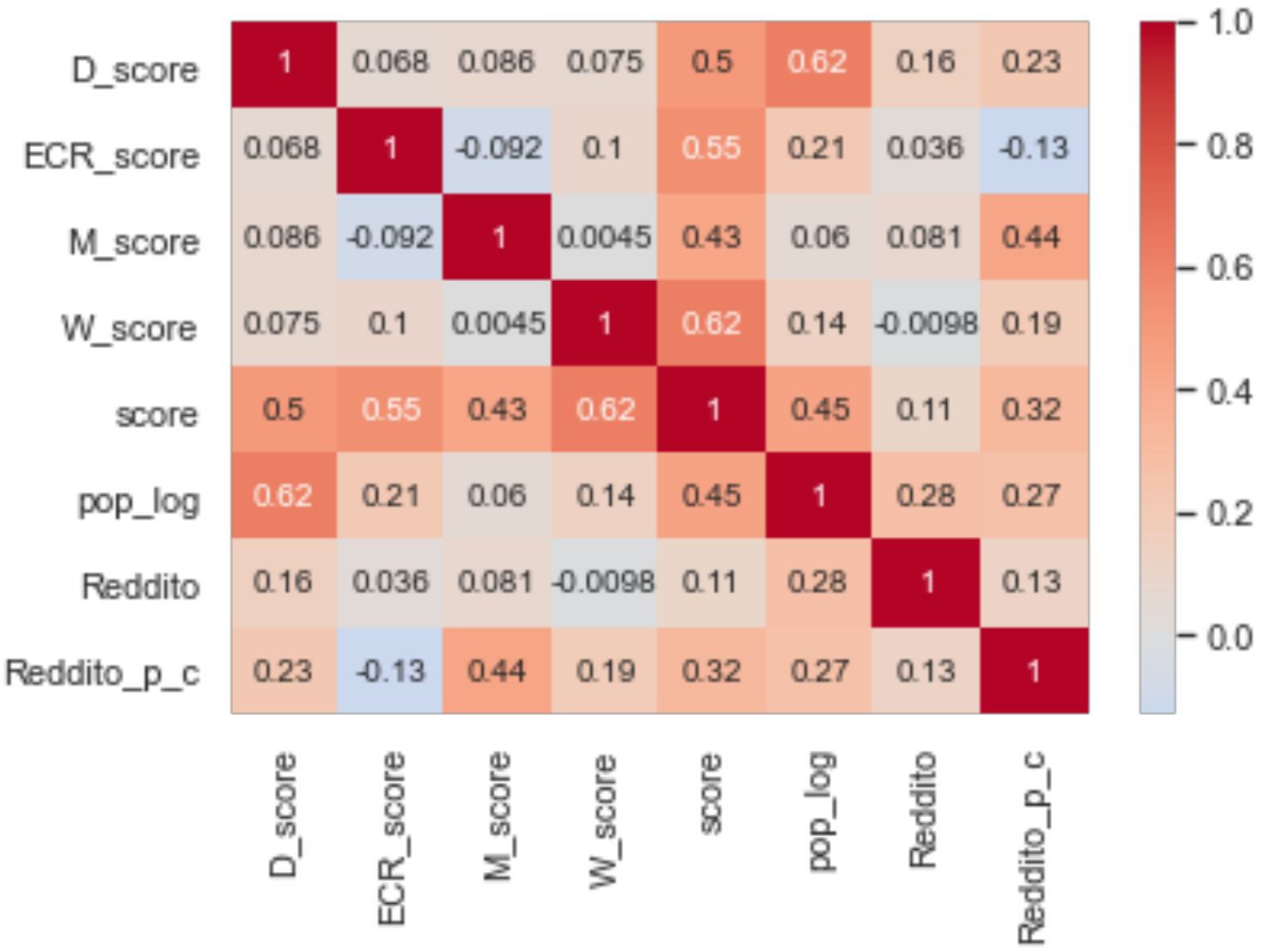

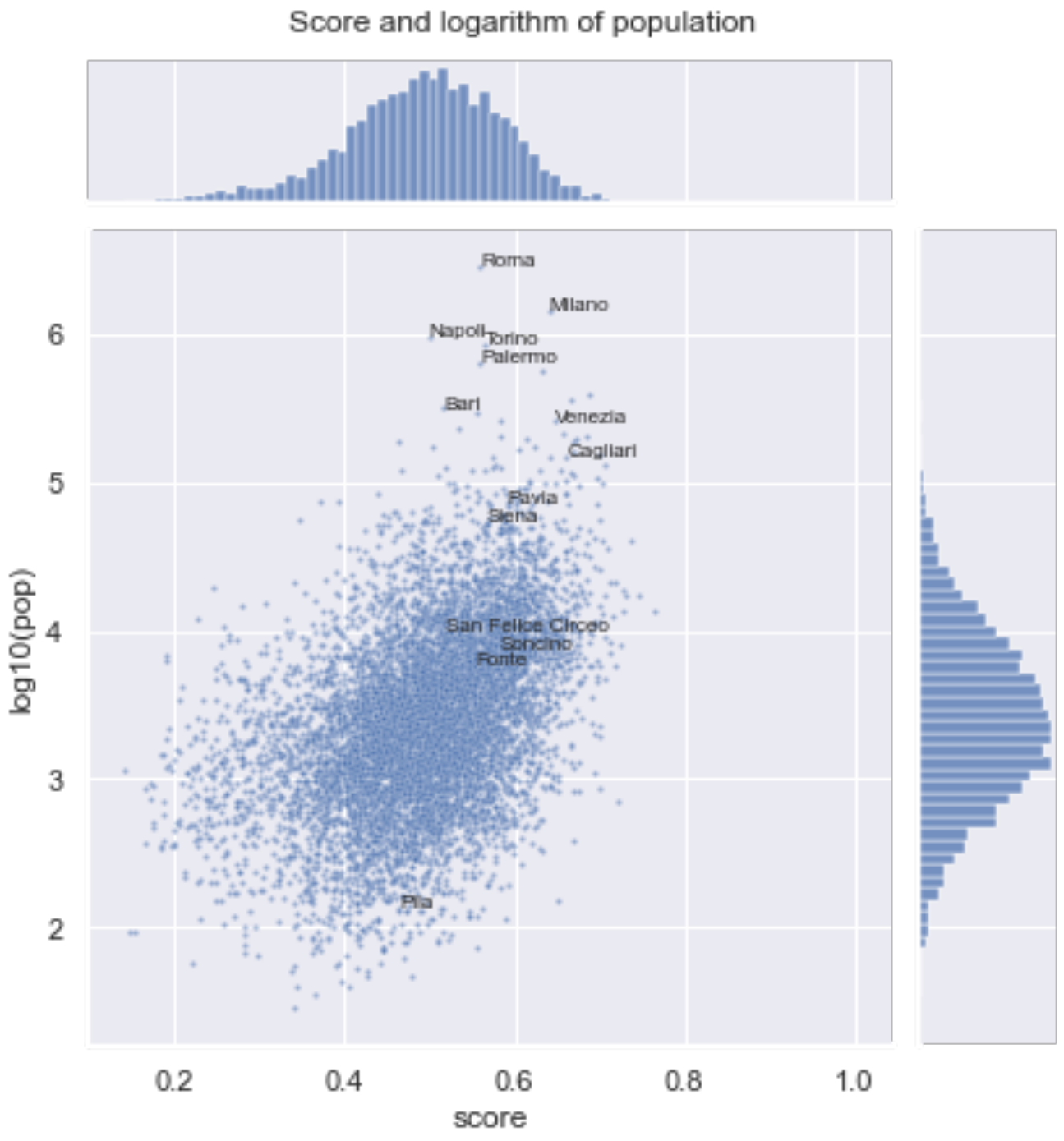

(b)

*Figure 9 (a) Correlation matrix of the main variables; (b) Detailed distribution of the logarithm of the population*

Considering the higher correlation values, Figure 9(b) shows the scatter plot and the population (log values) and CCI values distribution. The weak value of 0.52 confirms that the relationship is weak and probably due to nonlinear effects whose investigation is beyond the scope of this paper. Similar behaviour is seen in the other cases when correlations values exceed 0.4 (additional plots considering area values are in the Supplementary Information section).

*5.3 Sensitivity analysis and robustness check*

In this section, we test the robustness of the CCI with respect to the values of each area. According to the literature, area values are set to represent, with small variation, the importance of different areas in composing the final index value. Typically, area values show small differences and, in some cases, are set equal for each area (EIU, 2009; Dall'O, 2017; Lin, 2014). Because of this source of arbitrariness, we performed a set of robustness tests to understand how the final CCI distribution is altered according to area weights. The first test is made by comparing the CCI distribution discussed before to a baseline obtained by setting to the same area values (namely 0.25 each weight). Figure 10 shows the comparison between the CCI values distribution and the corresponding baseline. Excluding some small details, the two histograms are substantially equivalent, confirming the low sensitivity of the final distributions to slight variations in area weights.

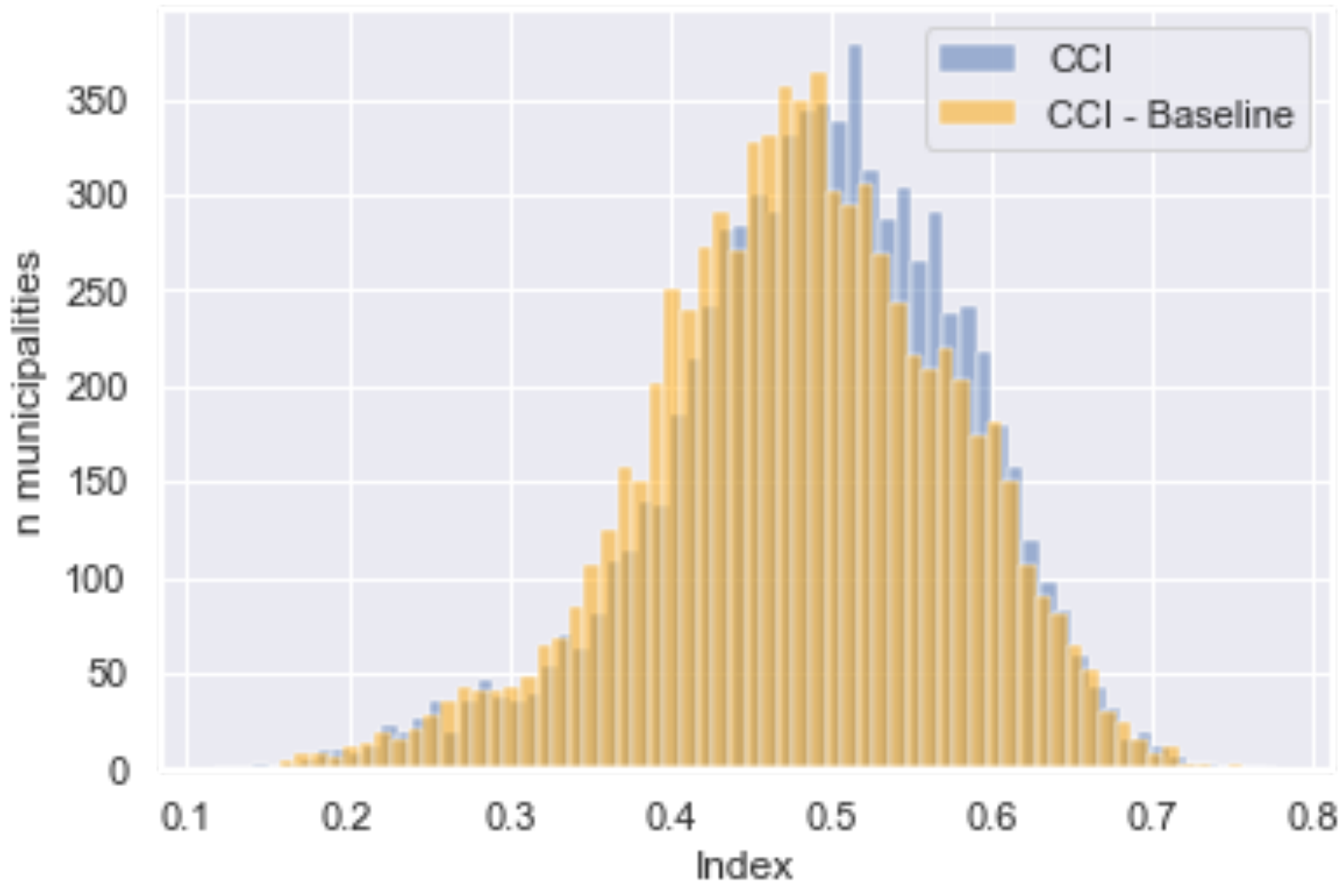

*Figure 10 Comparison between the selected area weights and the baseline weights*

The histogram also suggests that larger variations on the area weights are needed to understand how sensitive the area values in the presence of consistent variations. To this end, we performed a comparative statics analysis on the area weights: each weight value is varied in the interval [0.05,0.5], while to maintain the total area weight sum equal to 1, the values of the other area weights are adjusted consequently. By adopting this strategy, we can test at the same time the consistent variation of the area weight under analysis and the slight variation of the other weights. Figure 11 shows the change in the distribution of the CCI as the area weights vary over the interval [0.05,0.5]. As in the previous case, we do not observe significant variations in either the shape or the mean value of the CCI distribution for both D and ECR areas. M and W show slightly larger variations, especially for higher (but highly biasing and unrealistic values of the area weight, 0.05-0.15 and 0.4-0.5). The same is observed when the Waste is considered: the final value are higher if the W area is considered in the lower bracket of the areas weight variation. This additional detailed analysis confirms the robustness of the CCI even when there is consistent variations in the area weights.

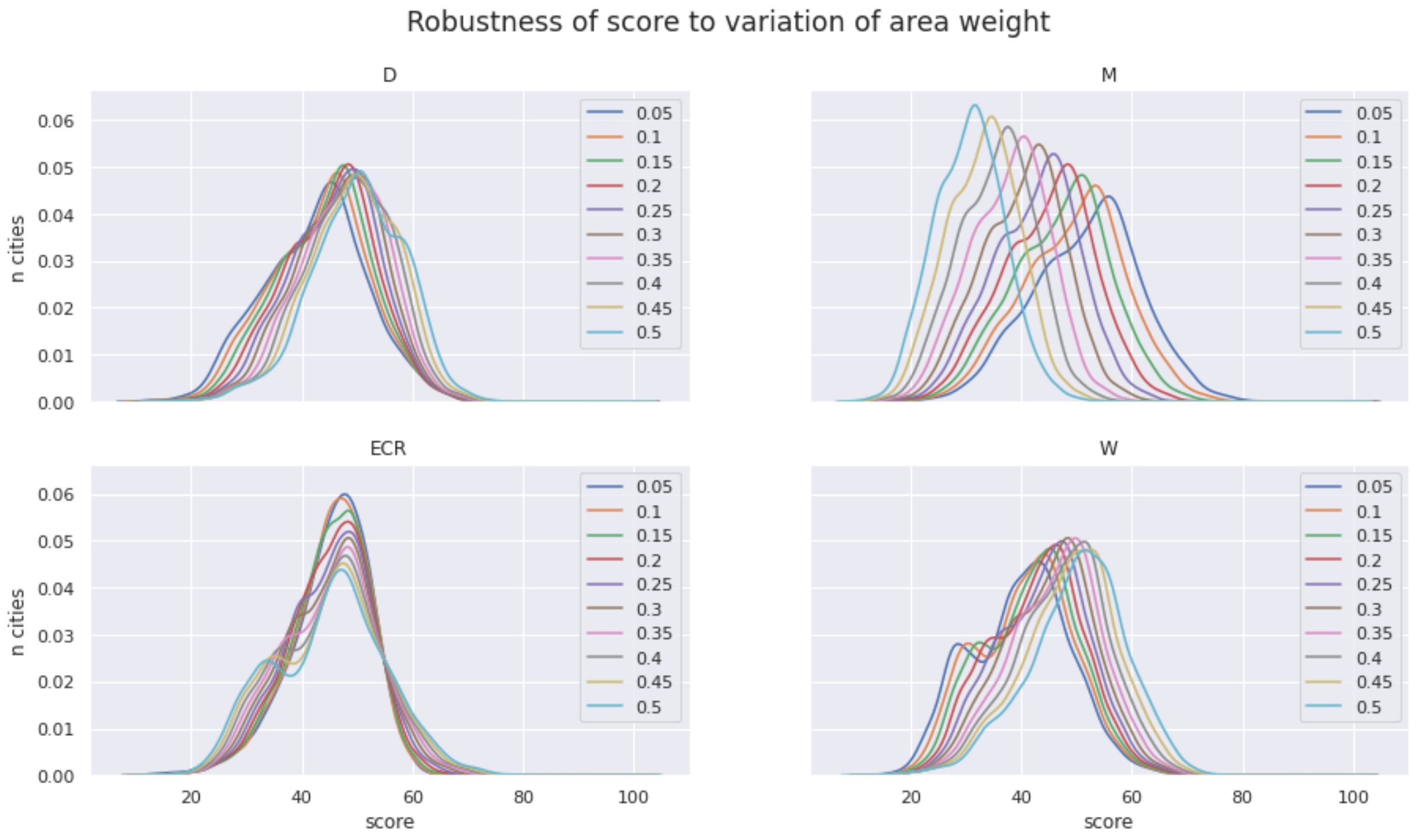

*Figure 11 Variation of the CCI distribution according to a single area weight variation in the range 0.01-0.5. D: Digitalization; M: Mobility; ECR: Energy, Climate, Resources; W: Waste*

From the geographic point of view, we are also interested in understanding whether regional patterns emerge when considering substantial changes in the index. To this aim, the map in Figure 12 shows the percentage variation experienced by a municipality CCI as the corresponding area weight vary. In general, the figures shows that most variations are on the order of 0.01-10%, while a small number of municipalities experience larger variations on the order of 10-30%. From the regional point of view, the most visible variations are found for Digitalization and Mobility in Southern Italy and partially in the islands for ECR and waste. It should be noted that, on average, the variations, even when considered in their upper bracket, remain absolutely limited to a maximum of 30% in a restricted number of cases. This is also confirmed by Figure 13, which shows the histogram of variations Δ. The average error is around 5-6% for each area, while only a limited number of municipalities show variations greater than 20%.Mobility is the only area that shows an average variation of 20%.

To summarize, the robustness check performed on the area weights reveals consistent robustness of the CCI even in the presence of values that empirically would not be accepted because their use can introduce biases in favor of a specific area. Therefore, the values selected in Table 3 appear to be consistent with the existing literature and robust considering the CCI distribution from both a statistical and geographic perspective.

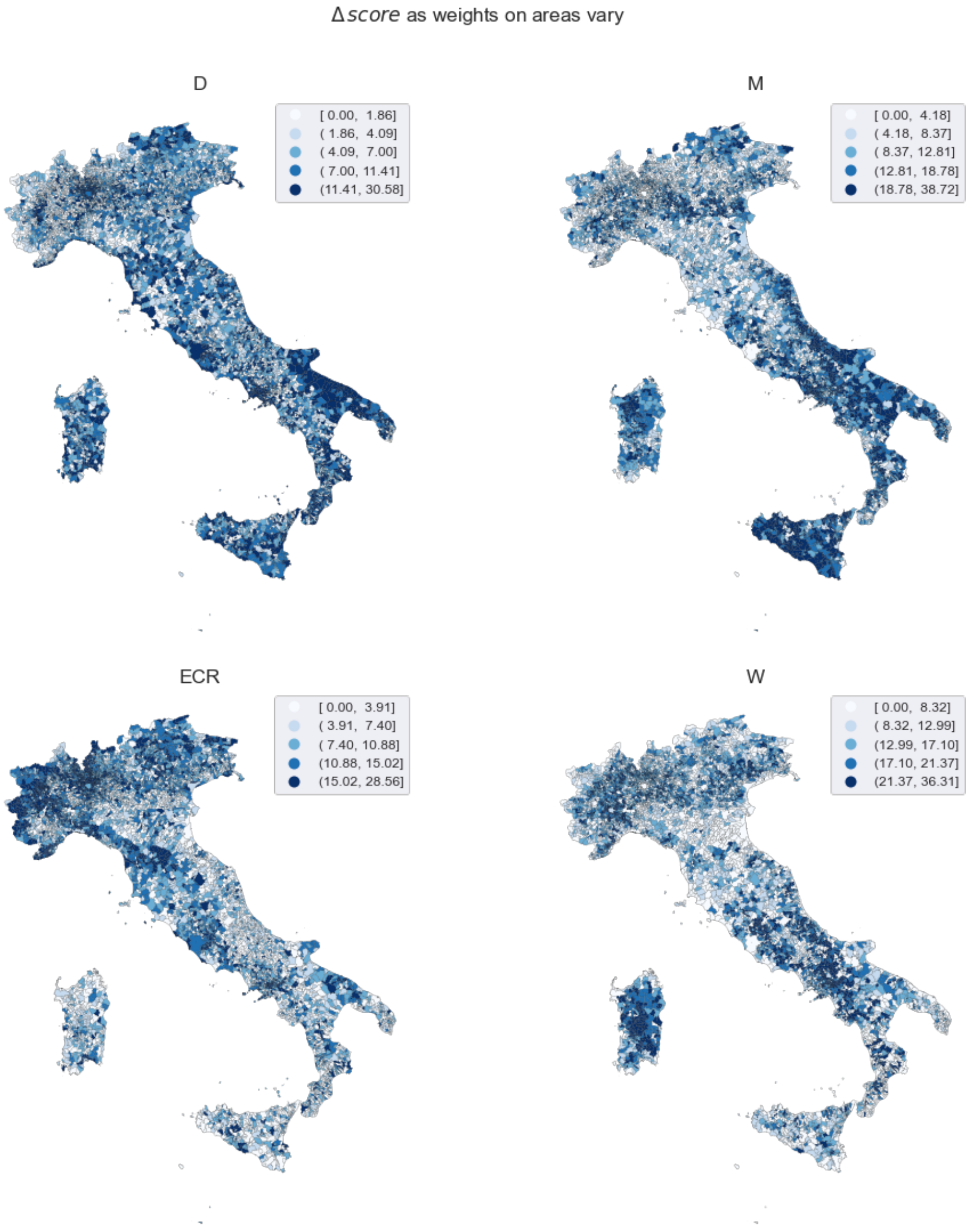

*Figure 12 Differences in Area values as the Area weights change.*

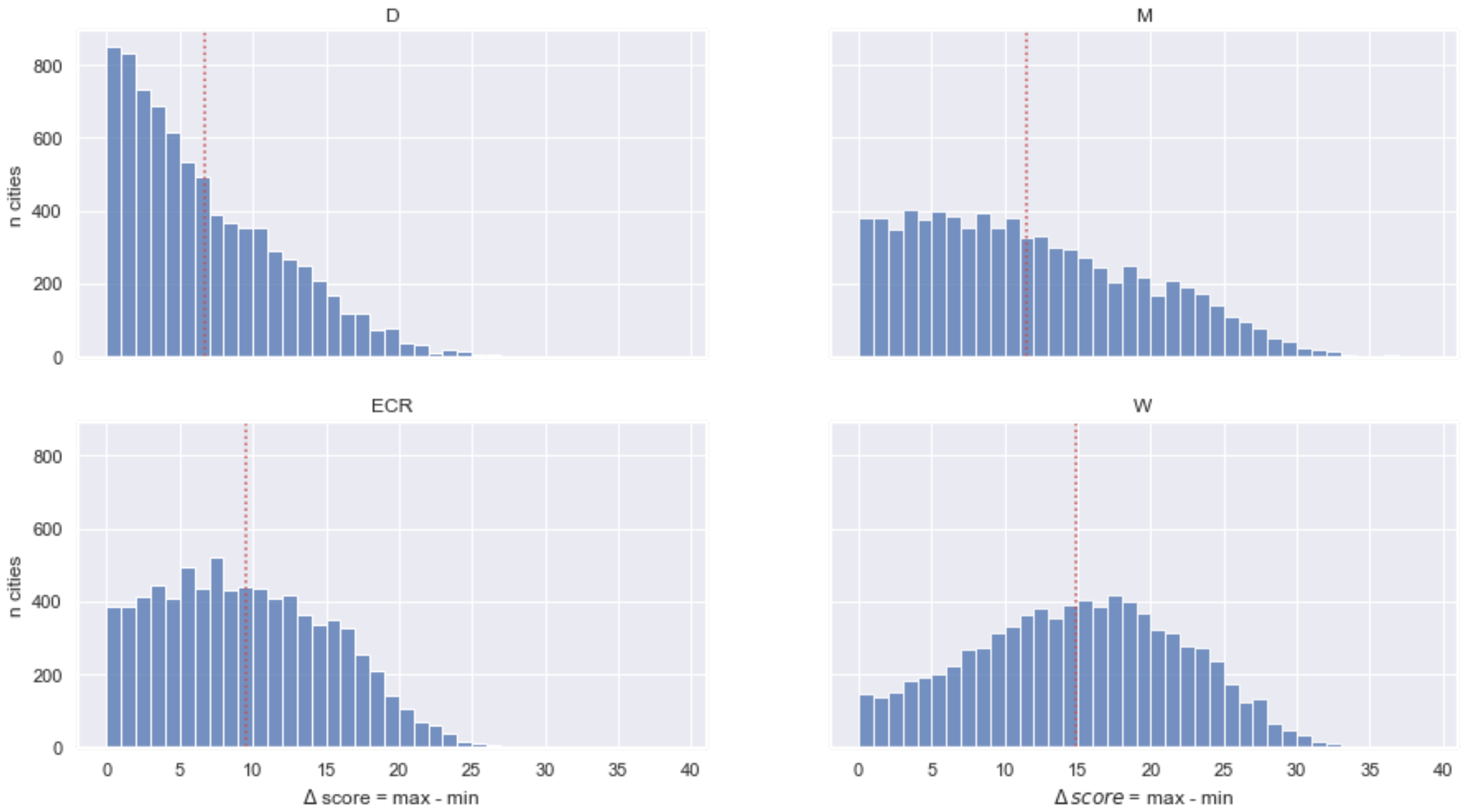

*Figure 13 Histograms of the maximum CCI difference for each area weight. The red line indicates the average of the distribution*

## 6. Discussion

In this paper, we propose the CCI index as a tool to evaluate the preparedness of the Italian municipalities in addressing the green transition, by providing a picture of how far cities are from the policy benchmarks established by the existing legislations. The CCI index provides also a synthetic description of the diverse situations at local level based on open data. The application of the CCI index to all the Italian municipalities blends convincing bias-free statistical characteristics with a normal CCI distribution. The ability to capture disparities both infrastructurally and regionally is also noted. The geographic analysis highlights regional disparities for both the overall CCI values and the four areas composing the index itself.

In particular, the area of Digitalization identifies a digital divide between large and small cities and across regions. Adoption of the ANPR digital platform is widely available among the municipalities, with only 300 lefts, accessibility to public digital portals and services is well represented, while broadband availability shows an important gap between rural and mountain areas compared to urban areas. At the same time, the north-south disparity does not seem to be significant for the Digitalization. In general, D is higher in large cities, but we also noted some unexpected exceptions. For example, Milan's D value is lower than that of other smaller cities (e.g., Livorno); this may be surprising because Milan is one of the municipalities where digital services are well developed and tested. Even so, the value is lower because of the small share of the population in possess of a broadband connection (about 60% versus over 90% of Livorno and other cities). In this case, the D values highlights its ability to identifies said gaps, and drive policy decisions towards a better distribution of infrastructure resources to reduce the population gap.

In the ECR area we note that, though a significant number of municipalities have signed the Covenant of Mayors[12], at the same time the commitment is generally low (i.e., a small fraction of municipalities committed for the 2030 targets and submitted an action plan), thus there is room for significant improvement. Municipal energy self-sufficiency shows high percentages in many municipalities (20% of the total reach 50% of self-sufficiency), with well-defined regional patterns that can be directly linked to renewable energy policies implemented in recent decades, especially in the islands and in the southern-eastern part of Italy. In the central-northern part, self-sufficiency seems to be lower, especially in the mountains and in the northwestern part. In this part, policies that encourage regional self-consumption are most needed to cover the gap. We also note that the self-sufficiency preparedness does not appear to be correlated with population size and income, while a portion of the municipalities showed self-sufficiency values above 100%, meaning a surplus of production at the territorial level. This is not always positive: under the infrastructural and economic perspective, this causes abnormal loads on the power grid that must be managed, leading to inefficiencies at DSO/TSO levels.

From the point of view of resources, we note that the average water loss of Italian infrastructure is generally high, with average values around 40%. Air pollution is also part of the ECR values and clearly reflects the presence of areas (especially the northern regions) where air quality is lowered by both industrial production and by meteorological factors. Because of the presence of weather factors in the present paper, we have assigned low weights to the related KPIs (ECR4, ECR5).

Mobility is the area where the disparity between northerners and the southern regions, as well as the disparity between large and small municipalities, is the greatest. This disparity can be traced back to the lack of coordination between policies and incentives encouraging light and sustainable mobility in recent years. An exception may be the lack of bike lanes, especially in municipalities located in mountain regions, which can now be remedied by the spread of e-bikes.

---

[12] Signatory cities pledge action to support implementation of the EU 40% greenhouse gas-reduction target by 2030 and the adoption of a joint approach to tackling mitigation and adaptation to climate change.

Looking at the distribution of the Waste area values, it may seem counterintuitive that southern regions show higher value with respect to northern regions. Sardinia island also shows a substantial number of municipalities with high values in waste transition preparedness. This can be explained by the construction of the area's own values that, by construction, recognizes those territories that shows both high shares of separate collection and low per capita waste production. This is motivated by the fact that, in our vision of circularity, transition should both increase the recycling and produce less waste. We also observe a decoupling between the waste production and both population and economic activity, as the waste does not appear to be correlated with both population size and income (per capita and total), as shown in Figure 9(a). An additional feature is the disparity between the capital city and surrounding municipalities. In Figure S1-10 we observe that, unlike the other areas and excluding some provinces, in general the surrounding municipalities show higher values. This may be explained by the fact that capital municipalities produce more waste and find it more difficult to increase waste recycling beyond the environmental standard set by law (65%). Other factors that may explain the results may be related to the territorial organization of waste collection. For example, the existence of a local collection policy may influence waste collection, and small municipalities can aggregate their collection and refer to another territory for municipal waste management and recycling.

### *6.1 Critical issues emerged*

Critical aspects emerged both from the perspective of data collection and analysis, and from the point of view of policy and environmental standards issues.

With regards to data collection and analysis, we point out the following issues:

1) *Lack of open sources* covering all Italian municipalities. Data are generally characterized by non-uniformity across years, outliers, and missing data. Moreover, depending on the year of release, they suffer from territorial non-uniformity, as Italian municipalities and provinces has been reorganized several times in the last decade. The lack of data affects aspects related to sustainability monitoring, excluding waste collection and air quality (although the latter concerns a small number of municipalities).

2) *Lack of environmental standards* to define adequate benchmarks. Excluding air quality and waste collection, the other KPI are not subject to rigorous environmental standards. Some benchmarks (see Table S1) have been set according to current policy recommendations, while the other, to the best of our knowledge, are not subject to recommendation. This may influence the scoring of those areas (e.g., mobility) where the presence of standard is low; in fact, the observed disparity may also be due to the voluntary adoption of policies to support the tools.

3) *Heterogeneity of KPIs.* Copying with different types of values and different range (e.g., binary, percentage, concentrations, etc.) poses problems in homogenizing the CCI calculation (see Supplementary information, Section S2, for further detail of the computation functions)

4) *Dimensionality reduction and linear combination of KPIs values.* Combining the values of individual areas into a single weighted index necessarily leads to a loss of information. Two municipalities with the same index may show differences in the areas and thus in the policies needed to implement the green transition. For a comprehensive analysis, the single area values should always be considered. Furthermore, the linear combination of the areas values into the final CCI could not account for potential nonlinear effects and nonlinear correlations that may be present in the data (e.g., non-negligible correlations between per capita income and waste)

## 7. Conclusions and future directions

In this paper we present a green transition preparedness index the Circularity City Index (CCI) that employs open data and measures and assess circularity of energy, material, and information. The index is, to the best of our knowledge, the first attempt to use open data to support green policies at national level considering a granularity down to the municipality level. We propose CCI as a policy tool, to better identify obstacles and structural problems, as well as a toll useful to define what can be the optimal transformation path for a municipality, taking into account the need of tailored policies that are aware of the specific potentials and criticalities of cities, town and urban and rural areas.

To compose the index, we identified a set of KPIs corresponding to factors that both directly and indirectly influence circularity and green transition, with a focus on the green new deal vision embraced by the European Union. The index is tested on the entire population of existing 7,904 Italian municipalities. The results show that the calculation of the index on a large national sample leads to a normal distribution of the CCI values, highlighting disparities both at the regional level and at the level of population size.

We also emphasize the strategic value of open data as a fundamental asset to define, plan and monitor the green transformation of municipalities on a national scale. Considering the Italian case, we propose the use of our index as a tool to better support a more efficient implementation of the PNRR recently issued by the Italian government. This may be particularly useful as we have observed that areas tend to offset others, so the enabling factors to trigger the transition may be different from one territory to another. Our index means to provide a preliminary tool to help the policymakers better understand enabling factors to promote a green transition as smooth and fair as possible for our cities.

Our future work will be devoted to a better characterization of the index by considering a broad spectrum of socio-economic variables, with analyses based on spatial econometrics methods to better define the effect of green transition policies, with possible relevant indication at both regional and local level. Finally, we believe that the development of clear and flexible tools will be critical to foster the dialogue and collaboration between scientists and policymakers in building together cutting-edge solution to foster the sustainable future of our cities.

**Acknowledgements**

The study is funded by EnelX. AM acknowledges funding from the Italian Ministry of Education "Progetti di Rilevante Interesse Nazionale" (PRIN) grant 2017ELHNNJ; AF is supported by SoBigData++ (Gr. N. 871042).

**References**

Brilhante, O., Klaas, J., 2018. Green city concept and a method to measure green city performance over time applied to fifty cities globally: Influence of GDP, population size and energy efficiency. *Sustainability,* 10.

CEC-COM, 2009. *GDP and beyond; measuring progress in a changing world.* Communication from the Commission to the Council and the European parliament, 433 final

Dall'O, *Green Planning for Cities and Communities*, Springer, 2020

Dall'O, G., Bruni, E., Panza, A., Sarto, L., Khayatian, F., 2017. Evaluation of cities' smartness by means of indicators for small and medium cities and communities: A methodology for Northern Italy. *Sustain. Cities Soc.* 34, 193–202.

Economist Intelligence Unit, *European Green city Index*, 2009

European Commission. *Making our cities attractive and sustainable. How the EU contributes to improving the urban environment*, 2010

Facchini, A., Kennedy, C., Stewart, I., Mele, R., 2017. The energy metabolism of megacities. *Appl. Energy* 186, 86–95.

R. Gehrig, B. Buchmann, Characterizing seasonal variations and spatial distribution of ambient PM10 and PM2.5 concentrations based on long-term Swiss monitoring data, *Atmospheric Environment*, Volume 37, Issue 19, 2003, Pages 2571-2580.

Gionis, A., Secondary, C.A., Author, C., Andrienko, G., Andrienko, N., Boldrini, C., Cintia, P., Cresci, S., Facchini, A., Caldarelli, G., Giannotti, F., Gionis, A., Tesconi, M., Guidotti, R., Mathioudakis, M., 2020. ( So ) Big Data and the transformation of the city. *Int. J. Data Sci. Anal.*

Jenks, George F. 1967. "The Data Model Concept in Statistical Mapping", *International Yearbook of Cartography* 7: 186-190

Likert R. (1932) Technique for the measure of attitudes *Arch. Psycho.*, Vol. 22 N. 140

Lin, J., Jacoby, J., Cui, S., Liu, Y., Lin, T., 2014. A model for developing a target integrated low carbon city indicator system: The case of Xiamen, China. *Ecol. Indic.* 40, 51–57.

G. Lucertini, F. Musco, Circular Urban Metabolism Framework, *One Earth*, Volume 2, Issue 2, 2020, Pages 138-142

Maranghi, S., Parisi, M.L., Facchini, A., Rubino, A., Kordas, O., Basosi, R., 2020. Integrating urban metabolism and life cycle assessment to analyze urban sustainability. *Ecol. Indic*. 112, 106074.

Organization for Economic Co-operation and Development. *How's life?: measuring well-being*. Paris.

G.M. Marcazzan, S.Vaccaro, G. Valli, R. Vecchi, Characterization of PM10 and PM2.5 particulate matter in the ambient air of Milan (Italy), *Atmospheric Environment*, Volume 35, Issue 27, 2001, Pages 4639-4650

Maurya, S.P., Singh, P.K., Ohri, A., Singh, R., 2020. Identification of indicators for sustainable urban water development planning. *Ecol. Indic.* 108, 105691.

A. Merino-Saum, P. Halla, V. Superti, A. Boesch, C.R. Binder, Indicators for urban sustainability: Key lessons from a systematic analysis of 67 measurement initiatives, *Ecological Indicators*, Volume 119, 2020, 106879.

Sáez, L., Heras-Saizarbitoria, I., Rodríguez-Núñez, E., 2020. Sustainable city rankings, benchmarking and indexes: Looking into the black box. *Sustain. Cities Soc.* 53

A. Sharifi, Urban sustainability assessment: An overview and bibliometric analysis, *Ecological Indicators*, Volume 121, 2021, 107102.

Stiglitz, J. E., Sen, A., & Fitoussi, J. P. (2010). *Mismeasuring our lives: Why GDP doesn't add up*. The New Press.

Taylor, Z. (2014). 'Lies, Damned Lies, and Statistics': *A Critical Examination of City Ranking Studies.* Intergovernmental Committee for Economic and Labour Force Development.

United Nations, *UN World Urbanization Prospect*, the 2018 revision, 2019

UNDP, *an Development Index* Technical Notes, 2020.

## Supplementary information

### S1. Data collection and treatment

Data have been collected according to the sources listed in SI Table 1. To design and feed the index a massive discovery on Italian Open Data landscape has been addressed, analyzing both Institutional Portal datasets as well as Open Data Communities. To minimize bias and to overcome the fragmentation currently affecting Italian Open Data scenario we've based our analysis on sources and datasets aligned to following parameters

a) Institutional/Certified sources or as alternative community sources already validated by National Institute of Statistics or Research Entities in European Projects (ie. OpenStreetmap, openchargemap)
b) Availability of records with granularity at the municipality level
c) Availability of records for most of the Italian municipalities, regardless their sizing and individual capability to publish open data
d) Availability of recent data – with updates in 2021 (best case) up to 2018 (worst case)

*S2.2 OpenStreetMap Features extraction*

Even though some Italian cities provide open data regarding their mobility infrastructures, these datasets are spotty and not equally updated so, since the project OpenStreetMap (OSM) provide a widespread number of features released as open data, and the Italian National Institute of Statistic (ISTAT) referred about these features as "a first coherent set of information for the different territories"[13], in order to get valuable features to define an objective mobility values at city level, we have identified the most suitable tag features from OpenStreetMap to spot mobility features on the following three main topic: pedestrian areas, cycleways and bus stop location.

[13] ISTAT is currently using OpenStreetMap as an experimental tool for the statistical analysis of road networks (https://www.istat.it/it/archivio/257382)

By taking the official OpenStreetMap Wiki as reference to identify the right tags (described in table 4), we extracted these features from OpenStreetMap by using the web-based data mining tool overpass-turbo.eu.

| **Pedestrian area** | **Cycleways** | **Bus stop** |
|---|---|---|
| place=square | cycleway=* | highway=bus_stop |
| highway=path | highway=cycleway | |
| highway=pedestrian | | |
| leisure=park | | |
| highway=footway together with foot=designated | | |

Each feature extracted from OpenStreetMap can be represented by one of these three types of geometries: polygon, line, or point.

By leveraging on this feature extraction and manipulation methodology, we were able to find the following highlights to validate the goodness of the OpenStreetMap source:

- Pedestrian
    1. 95% of the Italian municipalities have pedestrian features on OSM
    2. 47% of municipalities provide at least 9mq of pedestrian area for each in habitant
- Cycleway
    1. 62% of the Italian municipalities have cycleway features on OSM
    2. Over 20% of the Italian municipalities have less than 100m/km2 of cycleway
    3. Only 2% of the Italian municipalities have over 1,000m/km2 of cycleway
- Bus stop
    1. 57% of the Italian municipalities have bus stop features on OSM
    2. Over 125k bus stop features through the Italian territory

In order to avoid double counting caused by overlap geometries, overlapped geometries are firstly merged into a single polygon when they describe features related to the same topic, e.g., a pedestrian path that is fully located within a urban park, is merged into a single polygon because they both are pedestrian related. Once solved the overlap problem, we break up the geometries in overlap with the municipality shapes in order to get features relative to a single municipality. Then, to pair the features with his belonging city, we have calculated a representative point for each feature geometry and then paired this point with the municipality shapes to get in which municipality it takes place. With the aim of qualifying comparable quality index, we ultimately have calculated the following parameter:

pedestrian area per inhabitant, the cycleways length by the city area, and the number of bus stop for inhabitant.

**S2. Area values computation**

In this section we describe how the value $S_{kc}$ is computed starting from the KPIs. The first step is to define a benchmark (i.e. a reference value, an environmental standard, etc…) for each KPI, then, the value of the KPI is compared with the benchmark by means of a specific function that assigns for each KPI a value between 0 (benchmark not matched) and 1(benchmark fully matched). Because of the different types of KPIs (i.e. binary, number, etc) we distinguish the following types of functions:

1. **Binary**: $S_{kc} = kpi * 1$
2. **Percentage**: $S_{kc} = kpi * bench$ if *bench>0*, otherwise if *bench=0*, $S_{kc} = kpi - \min(K) / (\max(K) - \min(K))$, where K is the set of all KPIs collected for all the municipalities.
3. **Threshold_down**: This case covers the existence of an environmental standard with a threshold whose value should not be exceeded (e.g. air pollutant concentration). For this class of KPIs we designed a specific function mapping the value *kpi* into 5 intervals as follows: I1=[-1,0.5*bench]; I2=(0.5*bench,bench]; I3=(bench,1.5*bench]; I4=(1.5*bench,2*bench]; I5=(2*bench,\inf). Following this, we assign a score to each interval: I1=4, I2=3, I3=2, I4=1, I5=0. Finally, the values are normalized according to the normalization described in point 2.
4. **Threshold_up**: This case covers the existence of a standard with a threshold defining a minimum value (e.g. minimum number of charging stations). In this case we define $S_{kc} = kpi/bench$
5. **Quartile_down**: This case covers the case in which the production of pollutants or the use of a resource should be in the lower bracket. Using the quartiles of the KPIs values distribution cover the case in which a policy or an environmental standard is missing (e.g. per capita production of solid waste). Starting from the value distribution, we assign to each quartile a deacreasing score according to the following criterion: Q1=4, Q2=3, Q3=1,Q4=0. The resulting value is then normalized according to the formula described at point 2.

## S3. Additional tables and figures

*SI Table 1 Open source of KPIs*

| Level | Level weight | KPI | KPI weight | Definition | Data source | Benchmark source |
|---|---|---|---|---|---|---|
| Digitalization | 0.2 | D1 | 0.3 | Presence in ANPR (public digital service platform) | https://www.anpr.interno.it/ | Italian Digital Agenda[14] |
| | | D2 | 0.3 | Adoption of SPID in PA digital properties | https://www.spid.gov.it/servizi | Italian Digital Agenda |
| | | D3 | 0.3 | % of people with broadband connection (>30Mb/s) | https://maps.agcom.it/ | Italian Digital Agenda |
| | | D4 | 0.1 | Accessibility of local government digital properties | https://indicepa.gov.it/ipa-dati/dataset/servizi-digitali | Italian Digital Agenda |
| Energy, climate and resources | 0.3 | ECR 1 | 0.2 | Covenant of Mayors - Subscription | https://www.covenantofmayors.eu/en/ | See note[15] |
| | | ECR 2 | 0.2 | Covenant of Mayors - Level of commitment | https://www.covenantofmayors.eu/en/ | See note[16] |
| | | ECR 3 | 0.3 | % self-consumption (household only) | https://www.gse.it/dati-e-scenari/atlaimpianti | PNIEC, pag 56 |
| | | ECR 4 | 0.1 | Annual average concentration of PM10 | https://www.isprambiente.gov.it/ | D.Lgs. 155/2010 |
| | | ECR 5 | 0.1 | Annual average concentration of NOx | https://www.isprambiente.gov.it/ | D.Lgs. 155/2010 |
| | | ECR 6 | 0.1 | % of water leaks | http://asc.istat.it/ASC/ | See note[17] |
| Mobility | 0.2 | M1 | 0.2 | Pedestrian areas (m2/100 inhab.) | https://wiki.openstreetmap.org/wiki/Italy | Adapted: D.M. 2 aprile 1968, n. 1444 |
| | | M2 | 0.3 | Charging stations (points/1,000 inhab.) | https://wiki.openstreetmap.org/wiki/Italy | Art. 57, Legge 11 settembre 2020, n. 120 |
| | | M3 | 0.2 | Cycleways (km/100 km2) | https://wiki.openstreetmap.org/wiki/Italy | See note[18] |
| | | M4 | 0.3 | Bus stops (number/100 inhab.) | https://wiki.openstreetmap.org/wiki/Italy | Adapted from M2 |
| Waste | 0.3 | W1 | 0.4 | Per capita production of solid waste (t/inhab.) | https://www.isprambiente.gov.it/ | See note[19] |
| | | W2 | 0.4 | % of solid waste recycling | https://www.isprambiente.gov.it/ | EU directive 2012/19/UE |
| | | W3 | 0.2 | Collection of e-waste | https://www.isprambiente.gov.it/ | Decreto Legislativo 49/14 |

[14] http://presidenza.governo.it/GovernoInforma/documenti/piano_crescita_digitale.pdf
[15] Join of the Covenant of Mayors is on voluntary basis. In our opinion this step is strictly required to increase the Index of a municipality
[16] This tests the level of commitment of the municipality in embracing the values, aim and targets of the Covenant of Mayors. The higher the commitment, the higher the score
[17] Actually a strict environmental standard is missing, as the water utilities receive specific target for the reduction of water leaks in defined period of time (e.g., reduce leaks of 5% in two years). To be more strict, we set this benchark to 0%.
[18] No policy or recommendation is currently in force.
[19] No policy is currently in force

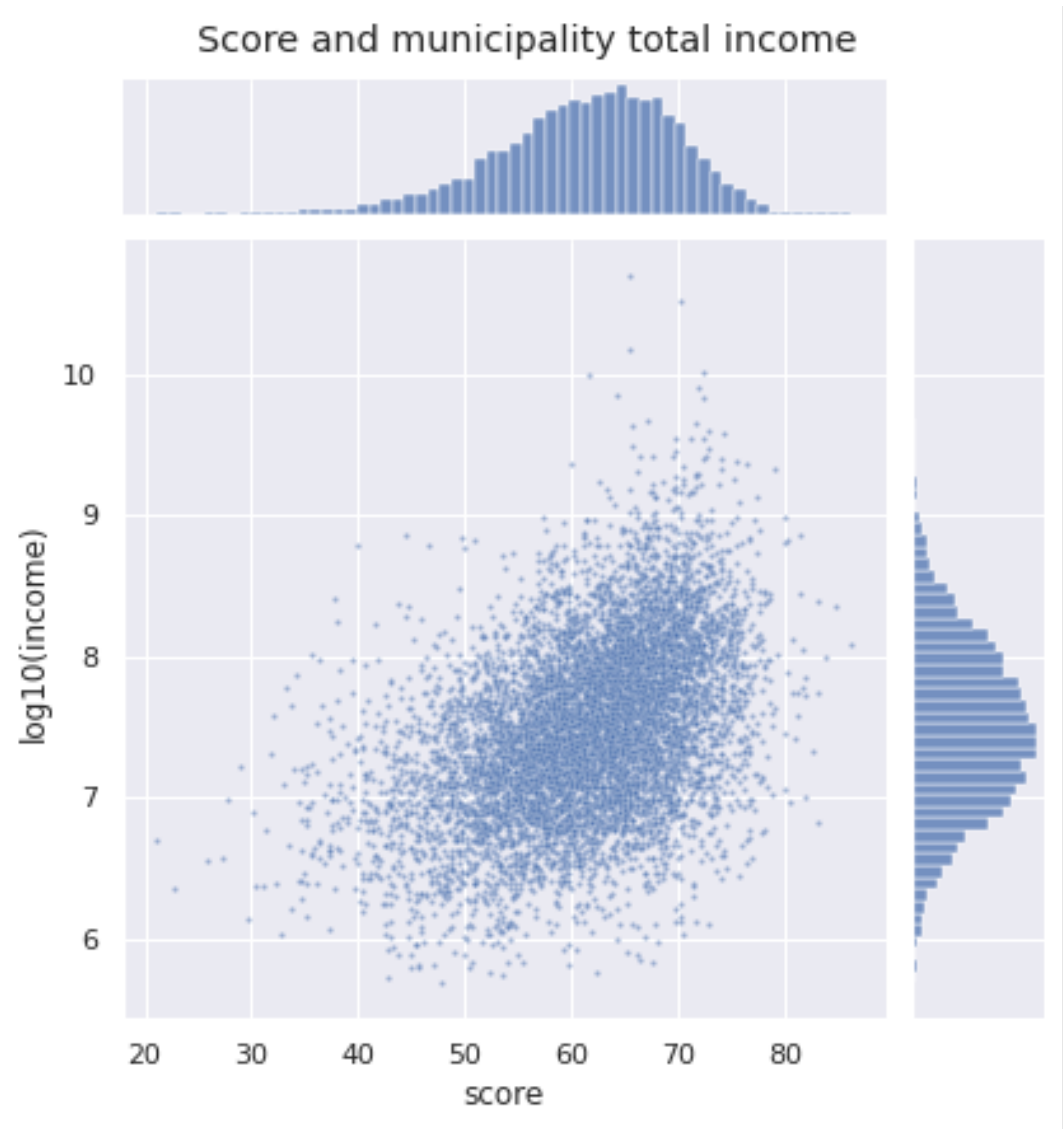

*SI Figure 1 Distributions of CCI and total income*

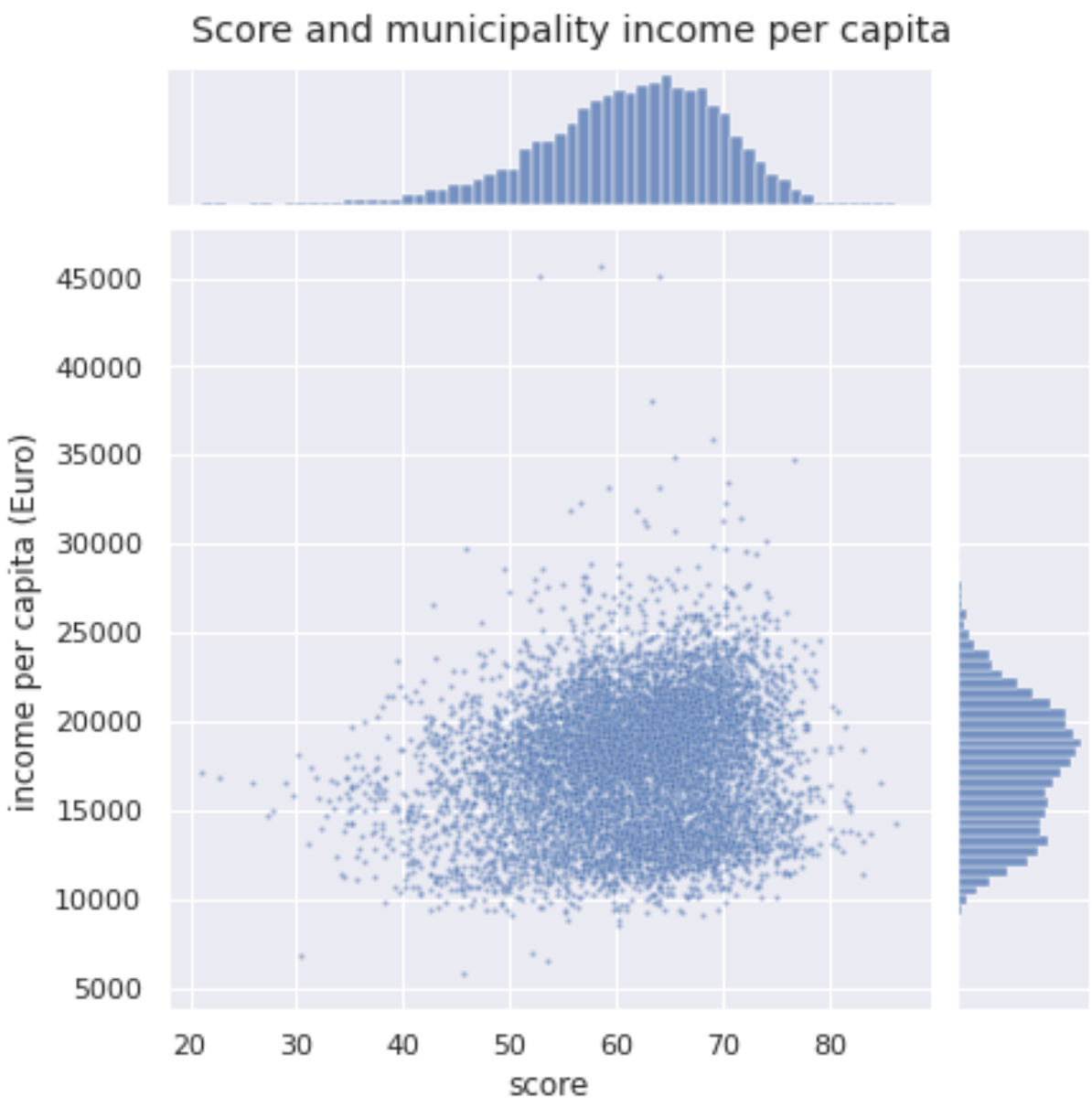

*SI Figure 2 CCI and income per capita*

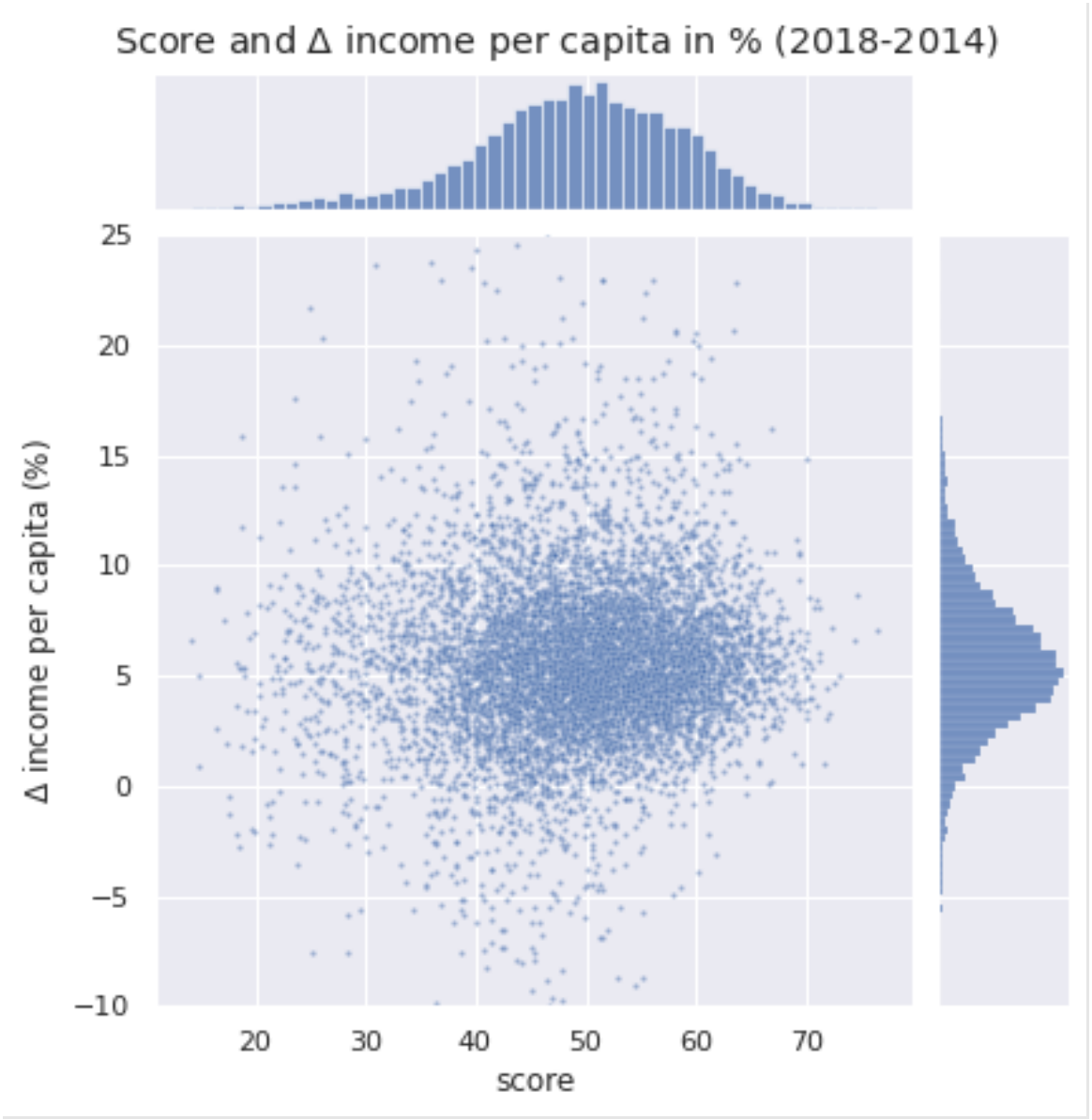

*SI Figure 3 Distributions of CCI and delta income (2014-2018) per capita*

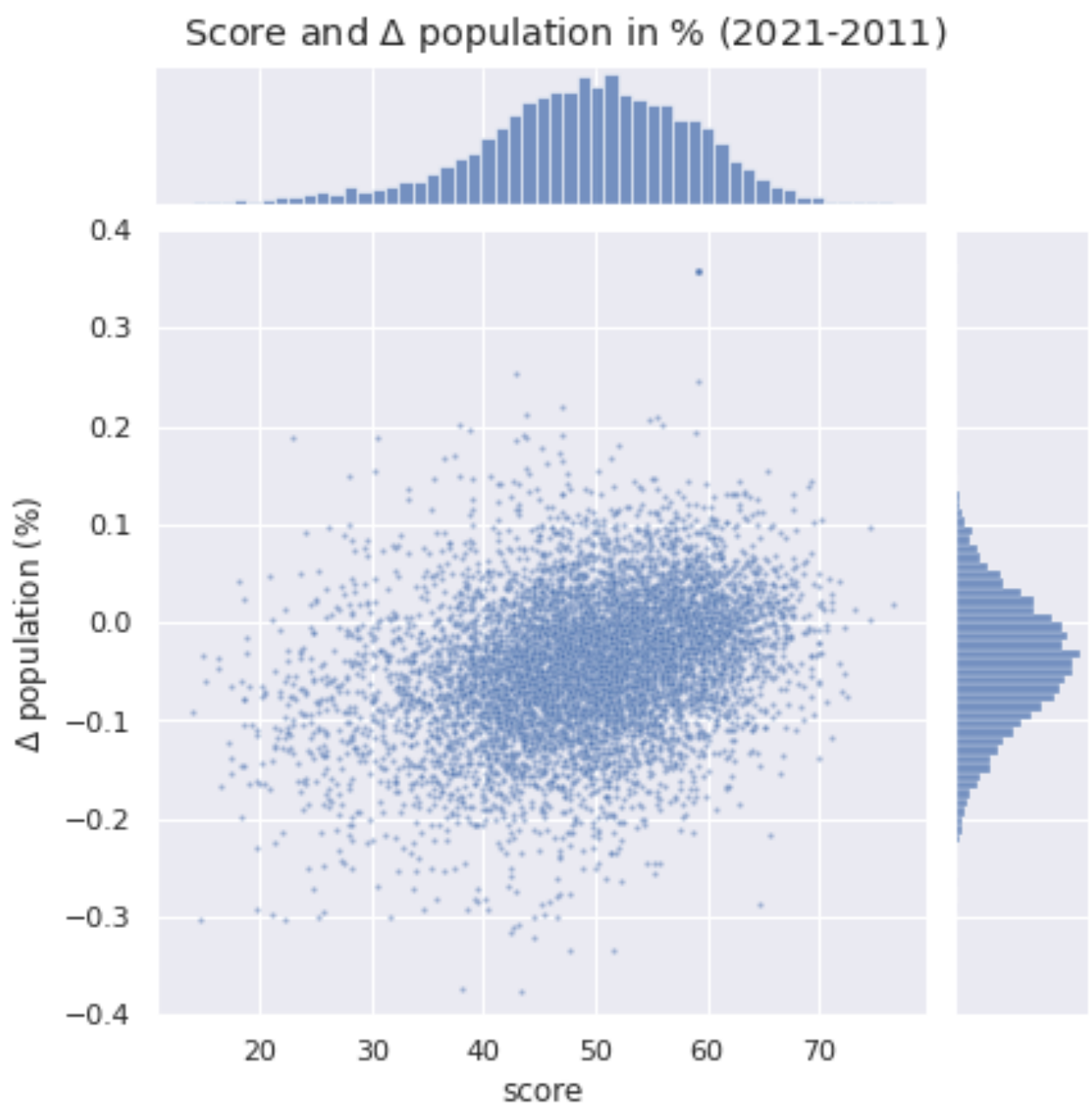

*SI Figure 4 Distributions of CCI and population variation (2021-2011)*

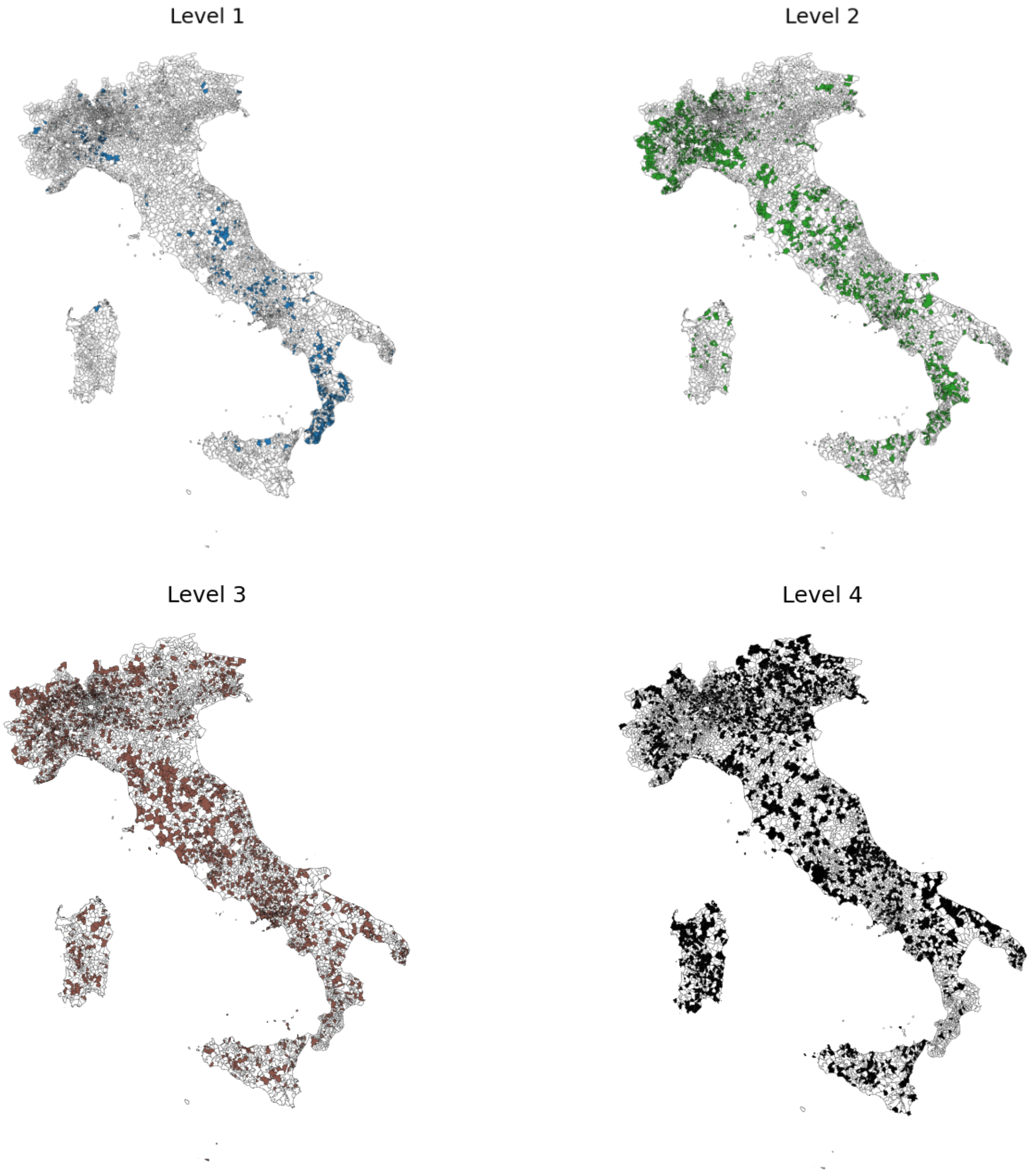
Level 1
Level 2
Level 3
Level 4

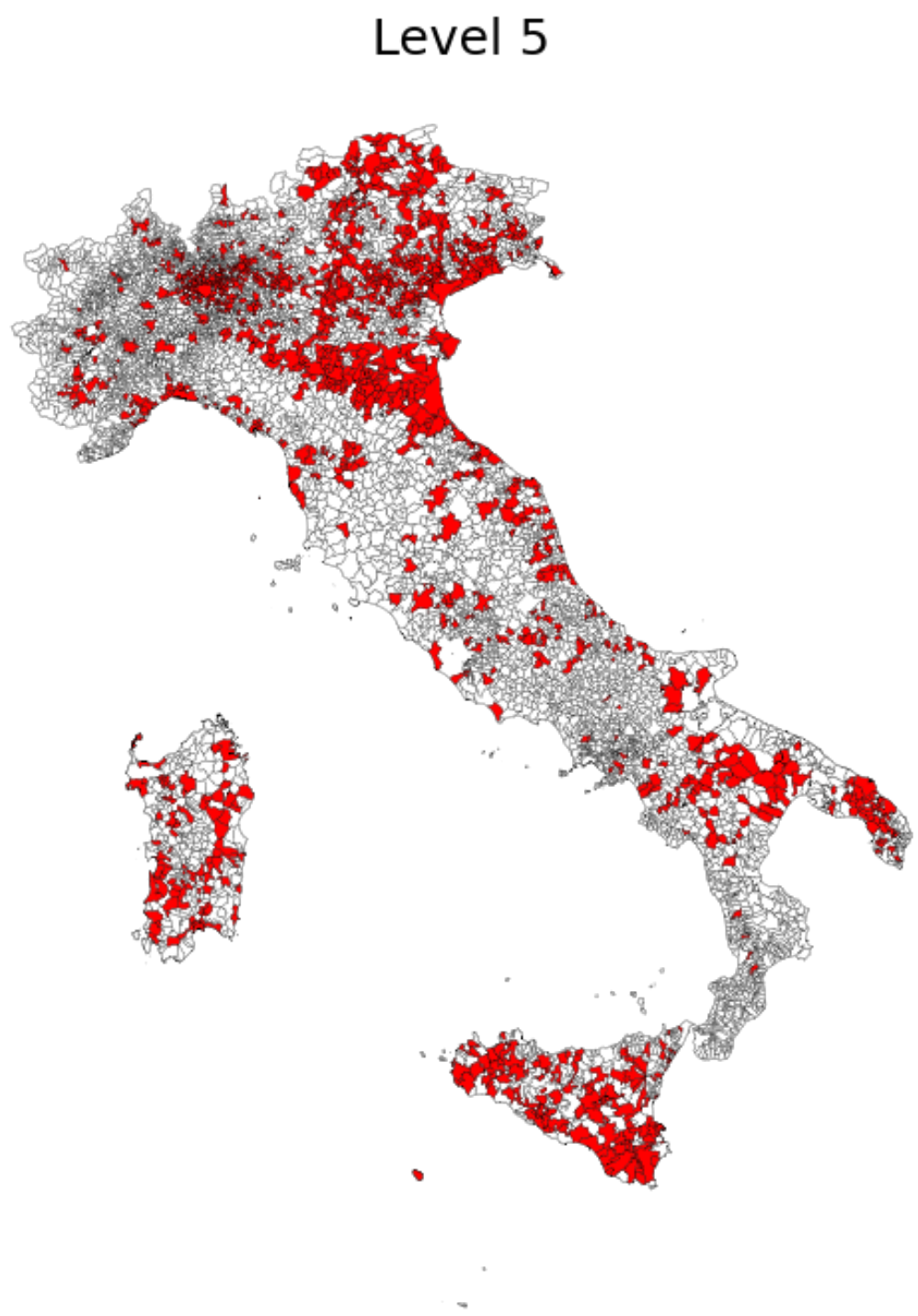

*SI Figure 5 Municipalities with CCI on the 5 levels of the Likert scale*

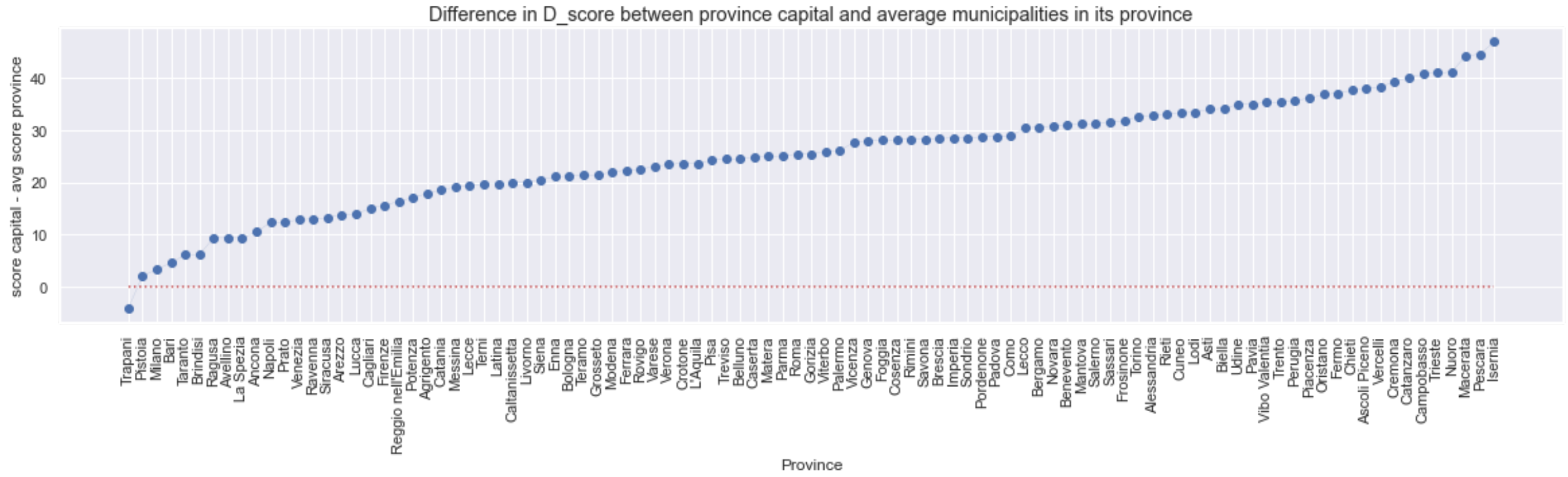

*SI Figure 6 Difference in D values between the average municipality and province capital*

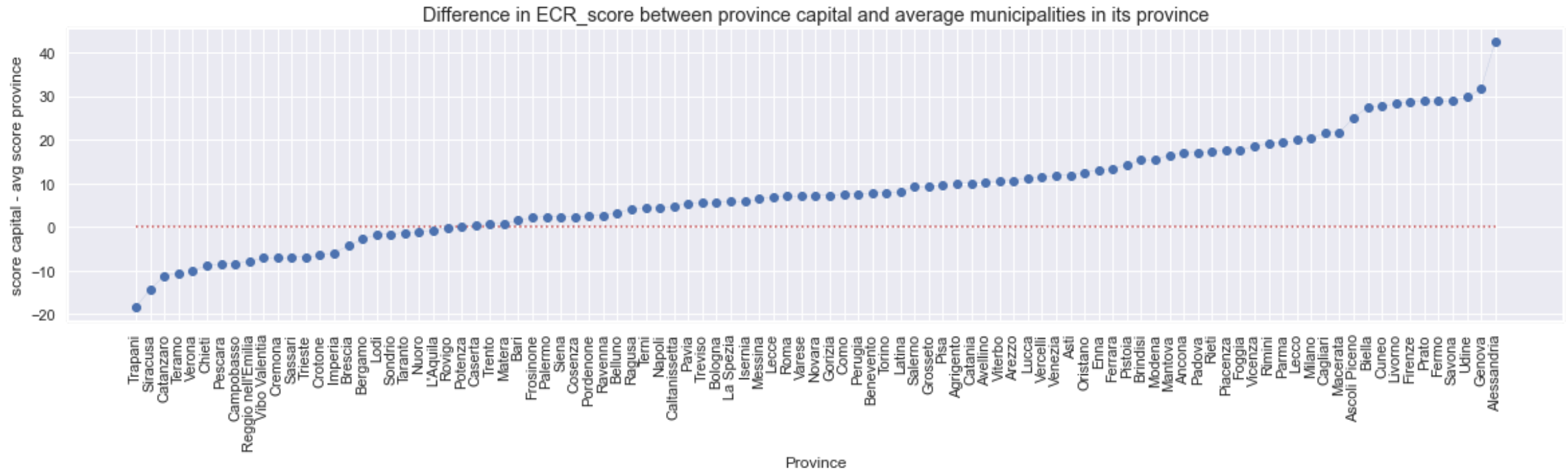

*SI Figure 7 Difference in ECR values between the average municipality and province capital*

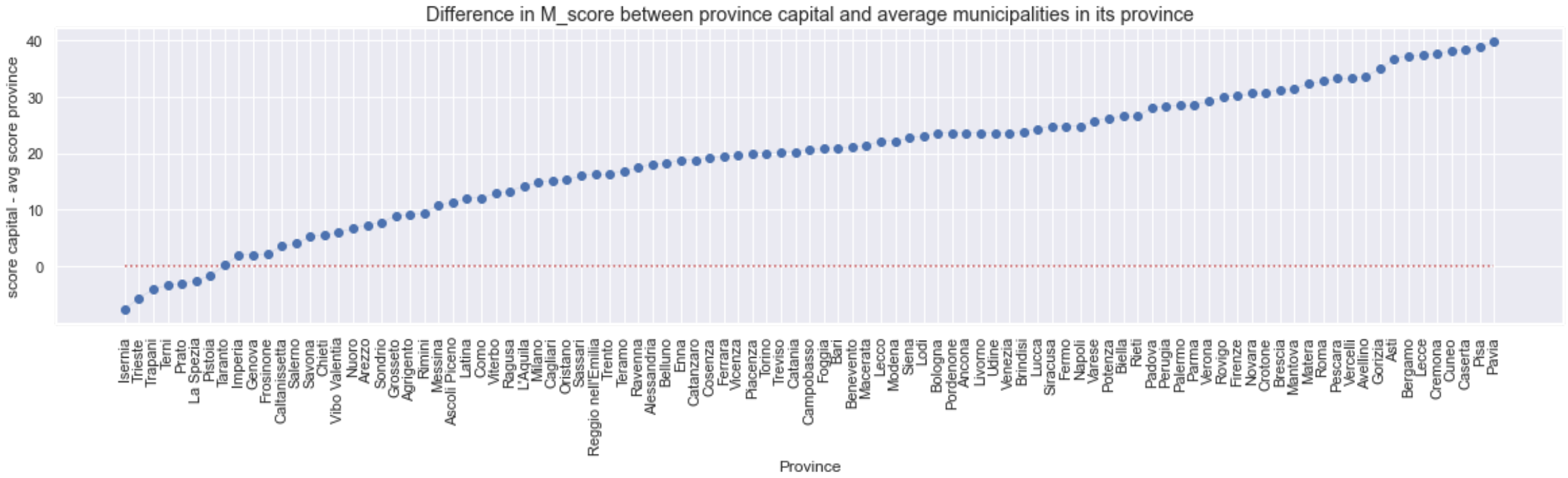

*SI Figure 8 Difference in M values between the average municipality and province capital*

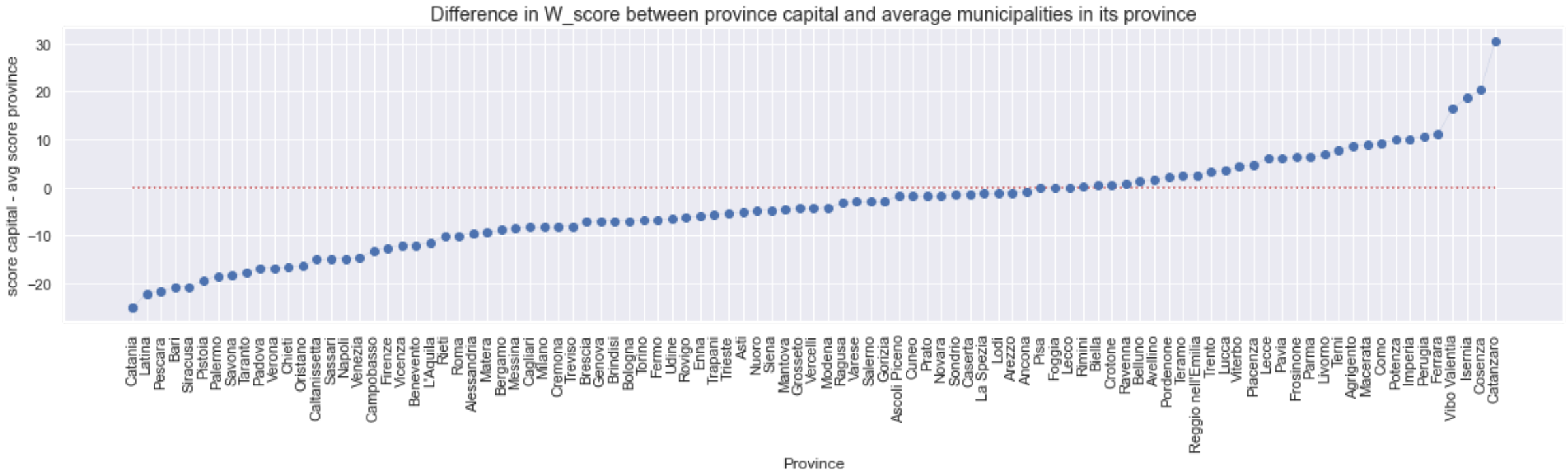

*SI Figure 9 Difference in W values between the average municipality and province capital*